\documentclass[conference]{IEEEtran}

\usepackage{url}
\usepackage{flushend}  
\usepackage{graphicx} 
\usepackage{times}    
\usepackage{url}      
\usepackage{cite}
\usepackage{balance}
\usepackage{color}
\usepackage{wrapfig}
\usepackage{caption}
\usepackage{subcaption}
\usepackage{array}
\usepackage{marginnote}
\usepackage{epstopdf}
\usepackage{xcolor}
\usepackage{mathtools}
\usepackage{amsfonts}
\usepackage{amsmath,amssymb}
\usepackage{epstopdf}
\usepackage{amsopn}
\usepackage{slashbox}

\usepackage{booktabs}

\usepackage{multirow}

\DeclareMathOperator*{\argmax}{arg\,max}

\def \sys {\textit{MagBoard}}
\def \asys {MagBoard}

\newcommand{\specialcell}[2][c]{%
  \begin{tabular}[#1]{@{}c@{}}#2\end{tabular}}

\begin{document}

\author{\IEEEauthorblockN{Heba Abdelnasser}
\IEEEauthorblockA{Wireless Research Center \\
Egypt-Japan Univ. of Sc. and Tech.\\
heba.abdelnasser@wrc-ejust.org}
\and
\IEEEauthorblockN{Moustafa Youssef}
\IEEEauthorblockA{Wireless Research Center\\
Egypt-Japan Univ. of Sc. and Tech.\\
moustafa.youssef@ejust.edu.eg}
\and
\IEEEauthorblockN{Khaled A. Harras}
\IEEEauthorblockA{Computer Science Department\\
Carnegie Mellon University\\
kharras@cs.cmu.edu}}

\title{MagBoard: Magnetic-based Ubiquitous Homomorphic Off-the-shelf Keyboard}

\maketitle

\begin{abstract}

One of the main methods for interacting with mobile devices today is the error-prone and inflexible touch-screen keyboard. This paper proposes \sys{}: a homomorphic ubiquitous keyboard for mobile devices. \sys{} allows application developers and users to design and print different custom keyboards for the same applications to fit different users' needs. 
The core idea is to leverage the triaxial magnetometer embedded in standard mobile phones to accurately localize a magnet within a virtual grid superimposed on a customizable printed keyboard. This result is achieved through a once in a lifetime fingerprint. \sys{} also provides a number of modules that allow it to cope with background magnetic noise, heterogeneous devices, different magnet shapes, sizes, and strengths, as well as changes in magnet polarity. Our implementation of \sys{} on Android phones with extensive evaluation in different scenarios demonstrates that it can achieve a key detection accuracy of more than 91\% for keys as small as $2\textrm{cm}\times2\textrm{cm}$, reaching 100\% for  4cm$\times$4cm keys. This accuracy is robust with different phones and magnets, highlighting \sys{}'s promise as a homomorphic ubiquitous keyboard for mobile devices.

\end{abstract}

\section{Introduction}

Mobile devices have become ubiquitous computing and communication devices. One of the main interaction methods with these devices is the touch-screen virtual keyboard. However, touch-screen keyboards have drawbacks including limited screen size causing user inconvenience and entry errors, occlusion of the keys by the user's fingers, fixed configuration options to support internationalization and/or special needs, and security issues that can leak users' passwords and other critical information \cite{wang2015mole}. Additionally, there are many cases where a physical keyboard can provide better experience for users ranging from kids to elderly with special needs.

Recently, a number of techniques have been proposed to replace the touch-screen keyboard including vision-based \cite{balzarotti2008clearshot}, acoustic \cite{zhu2014context,wang2014ubiquitous}, and wireless signals \cite{ali2015wikey}. Nevertheless, these techniques suffer from occlusion, interference from other moving humans, surrounding noises, and/or high energy consumption. In addition, they are only tailored to work with specific devices.

\begin{figure*}[!t]
\centering
        \begin{subfigure}[t]{0.3\textwidth}
                \centering
                \includegraphics[width=\textwidth]{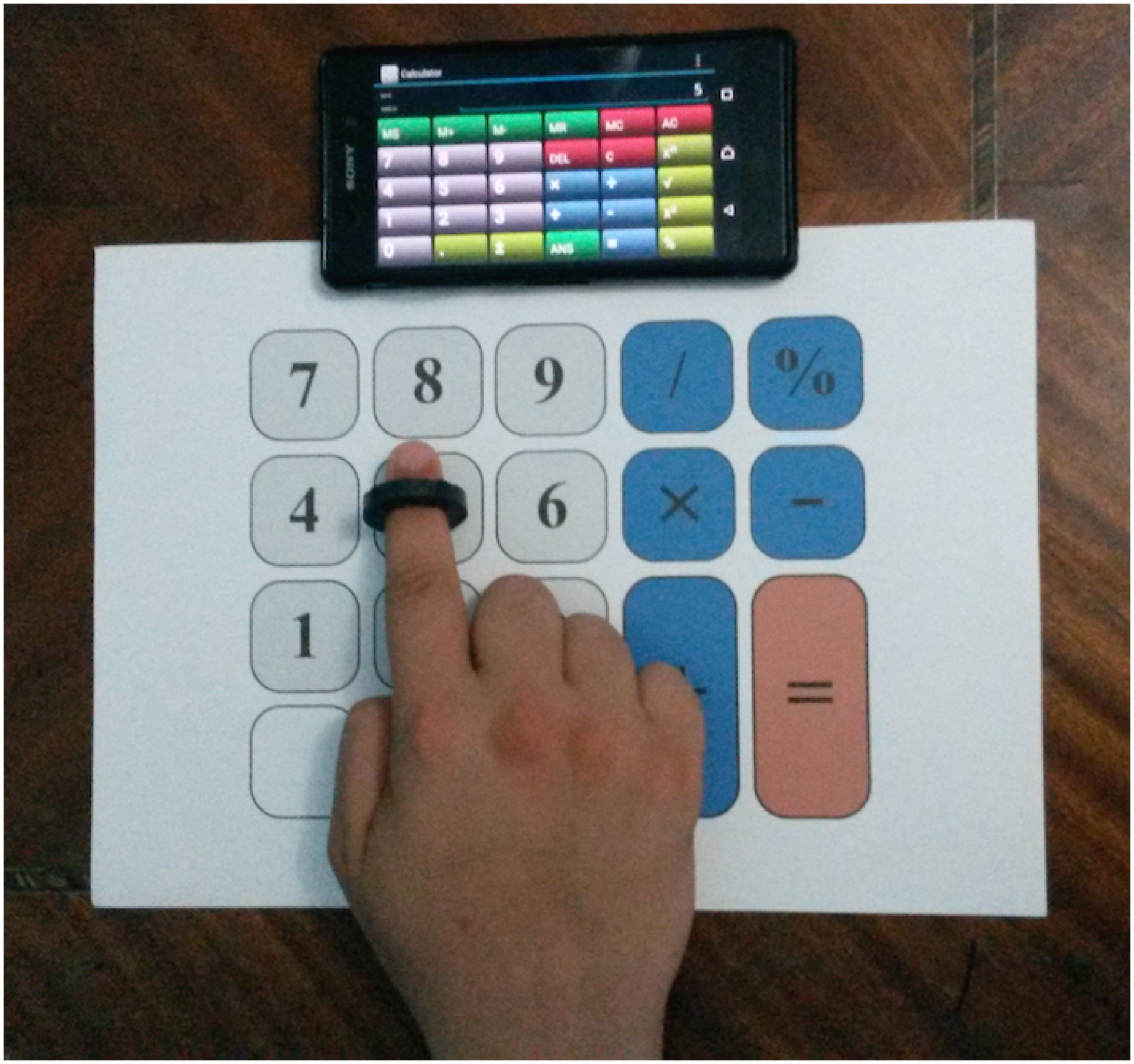}
                \caption{Larger keys.}
	      \label{fig:calc_app_intro}
        \end{subfigure}~~~~~~
        \begin{subfigure}[t]{0.3\textwidth}
                \centering
 	      \includegraphics[width=\textwidth]{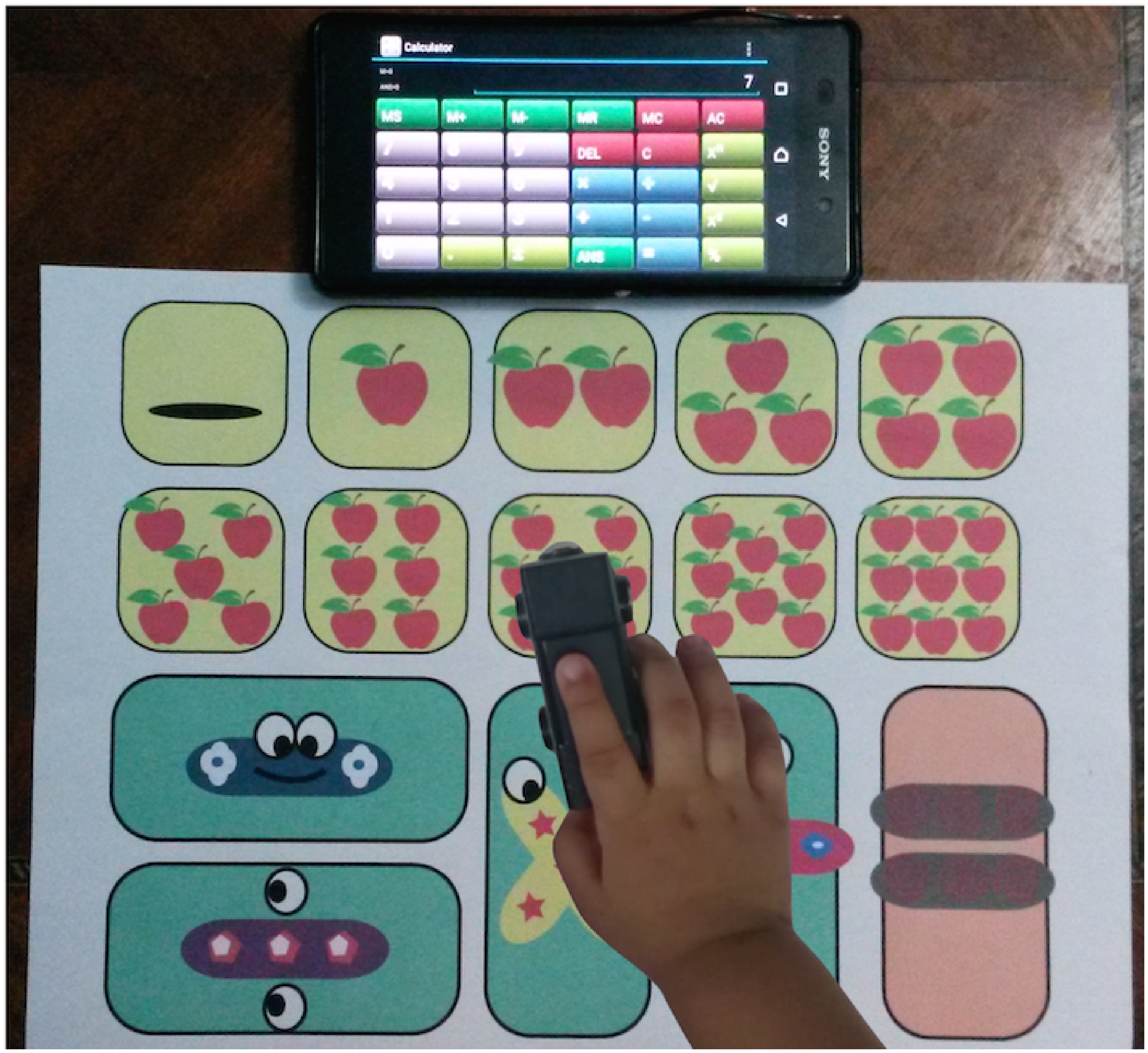}
               \caption{Kids-friendly keyboard.}
               \label{fig:calc_app_intro_kids}
        \end{subfigure}~~~~~~
        \begin{subfigure}[b]{0.3\textwidth}
	\includegraphics[width=\textwidth]{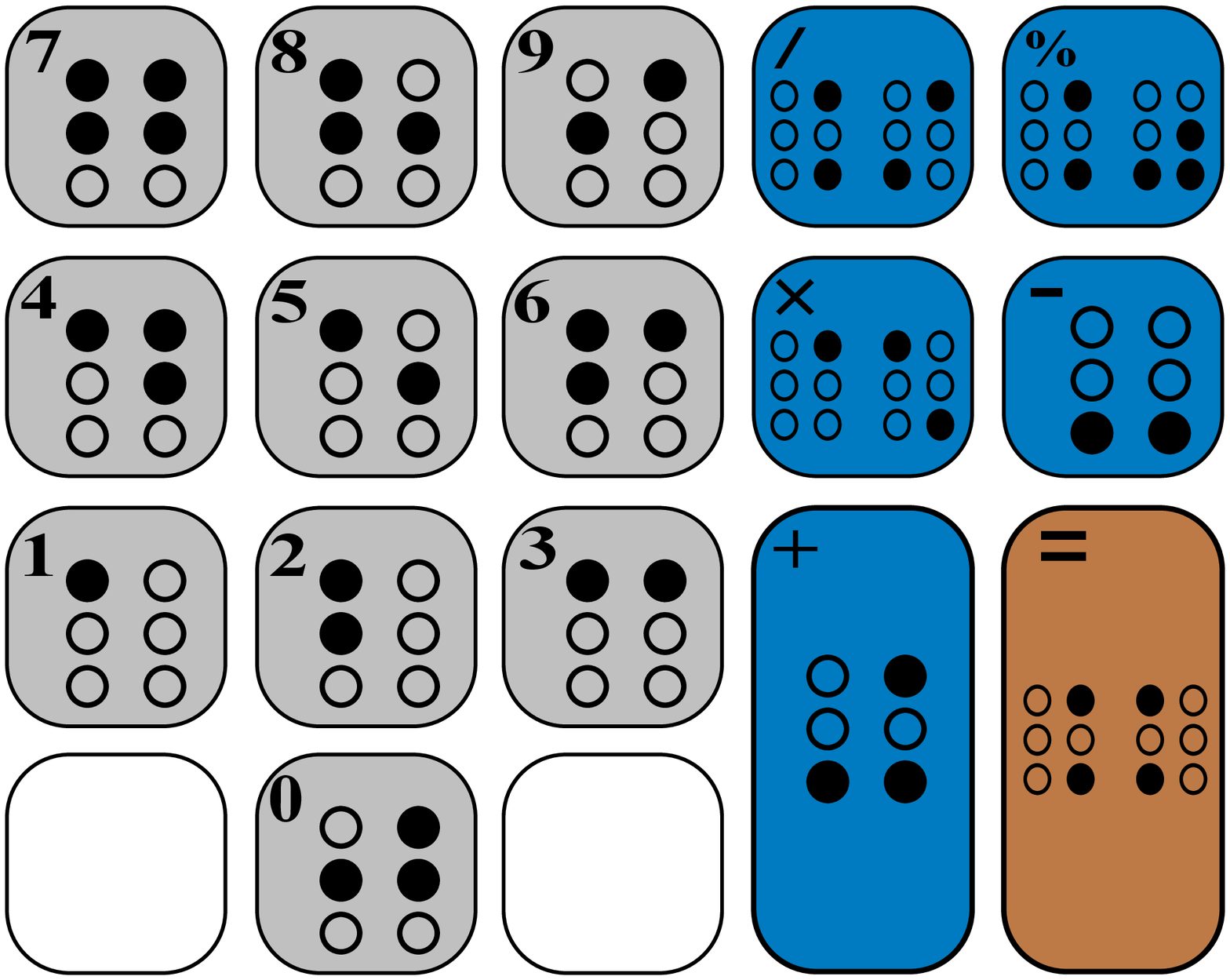}
	\caption{Calculator in brialle.}
	\label{fig:calculator_brialle}
	\end{subfigure}
\caption{Using \sys{} with a sample calculator application. Different physical keyboards can be used with the same application (homomorphism) to meet different users' needs.}
\label{fig:real_kids_calculator}
\end{figure*}

In this paper, we present \sys{}, a ubiquitous homomorphic off-the-shelf magnetic field-based keyboard for mobile phones. \sys{} allows the user to control the applications virtual keyboard using custom-printed keyboards (Figure~\ref{fig:real_kids_calculator}). The user utilizes an off-the-shelf magnet to specify the different keystrokes by determining the magnet location on a virtual grid imposed over the printed keyboard, leveraging the phone's built-in standard magnetometer. To accurately determine the magnet's location, \sys{} constructs a \textbf{\emph{one-time factory-based}} micro fingerprint of the virtual grid relative to the phone; during normal operation, it uses a probabilistic framework to achieve high key detection accuracy. \sys{} compensates for differences in phones, magnets, and polarity changes. It is also resilient to various environment noises through a \textbf{\emph{simple first-use process that takes less than one minute}}. Moreover, \sys{} provides application developers and regular users with tools to extend their applications with homomorphic keyboards (i.e. the same application can have different printed keyboards) that can fit varying user needs, which traditionally require OS changes or a lengthy error-prone software keyboard development process.

We deploy \sys{} on different Android phones and extensively evaluate its performance in different environments, with various orientations, users, and magnets. Our results show that \sys{} can achieve a key detection accuracy of more than 91\% under all scenarios. In addition, for a custom calculator keyboard application, it can achieve 100\% detection accuracy.

The remainder of this paper is organized as follows. Section~\ref{sec:system} presents the necessary background followed by the architecture and operation of \sys{}. Details of our implementation and extensive evaluation are shared in Section~\ref{sec:evaluation}. Section~\ref{sec:discussion} discusses different aspects of \sys{}. Finally, Sections \ref{related_work} and \ref{sec:conclusion} discuss related work and conclude the paper respectively.

\section{The \asys{} System}
\label{sec:system}

In this section, we present our \sys{} system. We first provide some background on the magnetic field and the notation that will be used in the remainder of this paper. Afterwards, we present the \sys{} system architecture along with a typical usage scenario. Finally, we describe in detail the different components that handle various challenges including variations in mobile devices, magnets, and users.

\subsection{Background and Notation}

\begin{table}[!t]
    \centering
    \begin{tabular}{|l|p{6.5cm}|} \hline
      \textbf{Notation} & \textbf{Description} \\ \hline \hline
      $B_T$ &  The total magnetic field vector.\\ \hline 
      $B_{\textrm{Earth}}$ &  The Earth geomagnetic field. \\ \hline
      $B_{\textrm{magnet}}$ & The magnetic field strength from the used magnet. \\ \hline
      $B_{\textrm{background}}$ & The magnetic field from the surrounding ferromagnetic materials.\\ \hline
      $B_{\textrm{silence}}$ & Magnetic field when the magnet is not present.\\ \hline
      $H$ & Hard-iron noise offset vector. \\ \hline
      $S$ & Soft-iron noise matrix. \\ \hline
      $B_p$ & Measured magnetic field by the phone's magnetometer.\\ \hline
      $\phi$ & Roll angle. \\ \hline
      $\theta$ & Pitch angle. \\ \hline
      $\psi$ & Yaw angle. \\ \hline
       $n$ & The number of magnetic field streams. \\ \hline
       $s$ & Magnetic field strength vector read by a magnetometer. \\ \hline
       $\mathbb{X}$ & The set of all virtual grid cells. \\ \hline
       $x$ & A cell in the virtual grid.\\ \hline
       $m$ & The number of successive samples. \\ \hline
       $k$ & Spatial averaging window size. \\ \hline

  \end{tabular}
    \caption{Notations used in the paper.}
    \label{tab:notations}
    \vspace{-0.2in}
\end{table}

To better understand \sys{}, we provide a brief overview on geomagnetic field basics and magnetometer operation on smartphones. All notations that will be utilized throughout the paper are defined in Table~\ref{tab:notations}. The triaxial magnetometer is a standard component of most current mobile devices. 
The sensor detects the cumulative magnetic field generated by multiple sources in the environment on each of its three sensor axes. The total magnetic field vector at any location ($B_T$) is the superposition of: the Earth's geomagnetic field ($B_{\textrm{Earth}}$), the magnetic field from the environment, which includes background magnetic field from the ferromagnetic materials ($B_{\textrm{background}}$), and the magnetic field strength from the \sys{}'s magnet ($B_{\textrm{magnet}}$). Therefore:

\begin{equation}
B_T = B_{\textrm{Earth}} + B_{\textrm{background}} + B_{\textrm{magnet}} 
\end{equation}

The magnetic field \emph{measured} by the phone's magnetometer, $B_p$, can be obtained from $B_T$ while taking into account the phone's three rotation angles: yaw ($\psi$), pitch ($\theta$) and roll ($\phi$):

\begin{equation}
B_p = R_x(\phi)R_y(\theta)Rz(\psi) B_T
\label{eq:Bt_wonoise}
\end{equation}

where  $R_x(\phi)$, $R_y(\theta)$, $Rz(\psi)$ are the corresponding rotation matrices. However, due to environment noise, hard-iron ($H$) and soft-iron ($S$) effects impact phone readings of $B_t$ \cite{ozyagcilar2012calibrating} as follows: 

\begin{equation}
B_p = S R_x(\phi)R_y(\theta)Rz(\psi) B_T + H
\end{equation}

Where the hard-iron effect $H$ is an offset vector and the soft-iron effect $S$ is a matrix. The effect of such noise on the magnetometer readings is to make the locus of the magnetic readings when the phone is rotated at a fixed point an ellipsoid rather than a sphere. The hard and soft-iron effects can be removed by the standard magnetometer calibration process performed by all mobile operating systems. Therefore, for the remainder of the paper, we focus on handling the components of Equation~\ref{eq:Bt_wonoise} so that \sys{} can work independently of the phone or magnet used, and regardless of the phone orientation.

\subsection{System Overview and Scenario}

\begin{figure}[!t]
\centering
\includegraphics[width=3.4in]{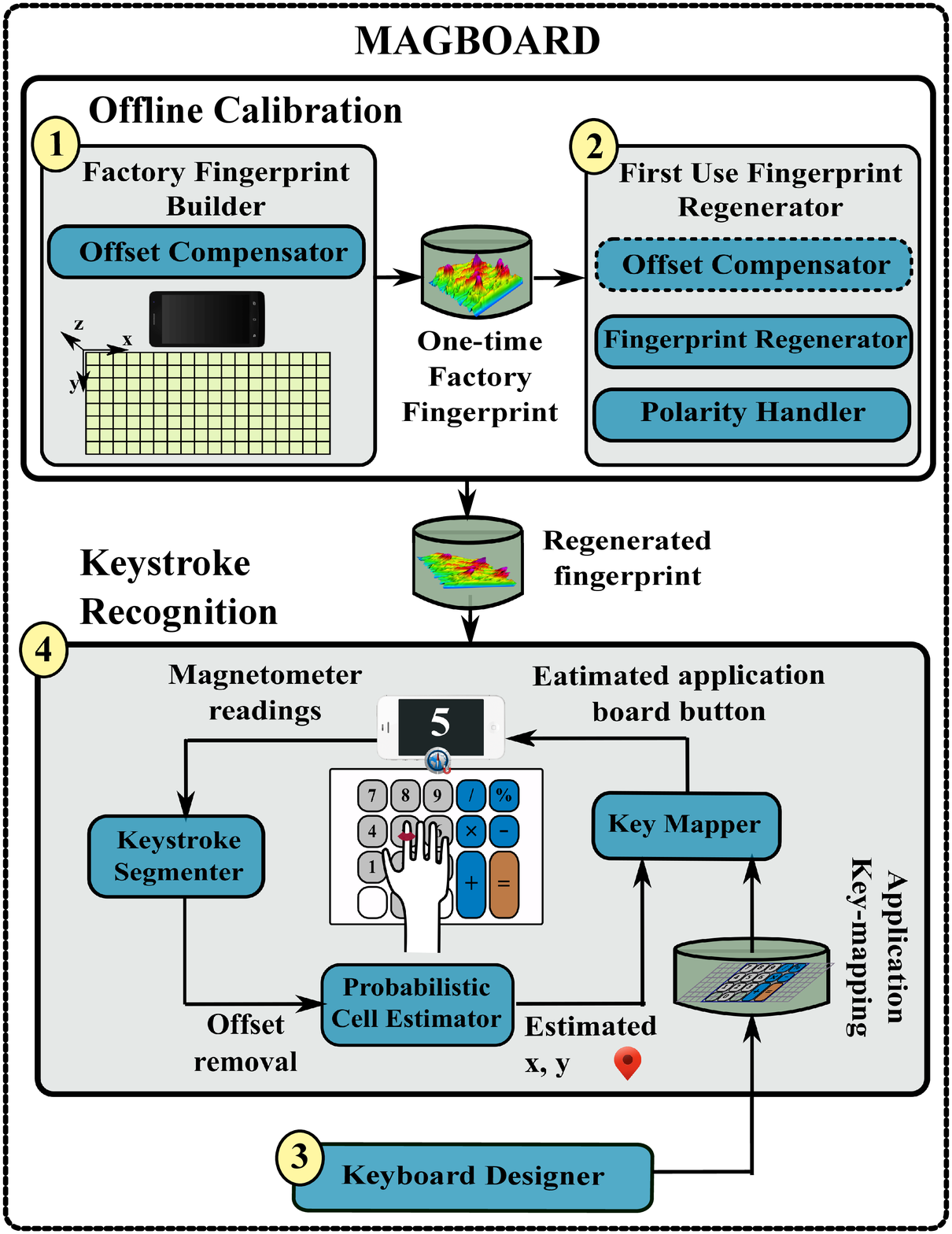}
\caption{\sys{} architecture. The system has four main blocks: 1) The once-in-life-time factory fingerprint building, 2) the first-time-use fingerprint regeneration, 3) the keyboard designer, 4) and the keystroke recognition.}
\label{fig:architecture}
\vspace{-0.2in}
\end{figure}

\begin{figure}[!t]
\centering
\includegraphics[width=3in]{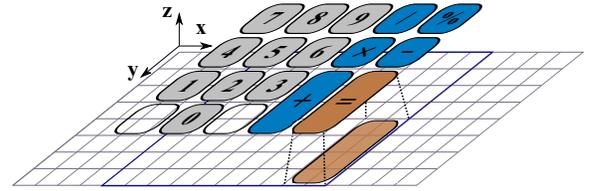}
\caption{Application grid mapping to the fine-grained virtual grid.}
\label{fig:app_grid_mapping}
\end{figure}

Figure~\ref{fig:architecture} shows \sys{}'s system architecture and typical operational scenario. To provide accurate magnetic-based keystroke detection, \sys{} relies on building a fingerprint of a virtual uniform grid relative to the smart phone. Note that causing a slight change or vibration in the phone, while typing for instance, does not cause the magnetic field distribution to change significantly. A cell in this virtual grid represents a single key of a keyboard at the \textbf{\emph{finest granularity}}. Application developers can use the ``\emph{Keyboard Designer}'' tool to group these cells into coarser-grained cells that map to the actual application keys (Figure~\ref{fig:app_grid_mapping}). This capability can be used to achieve different application goals such as having bigger keys, non-traditional keyboard layouts (e.g. curved keys), among others (Figure~\ref{fig:real_kids_calculator}). The resulting mapping is stored in an \textbf{\emph{Application-specific Key-mapping Database}}. The user can download and print a specific application layout as well as design and print her own customized keyboard layouts through the same module.

To reduce the calibration overhead, \sys{} builds the fine-grained fingerprint \textbf{\emph{only one time}}, independent from any users, devices, or magnets (we call this \textbf{\emph{``one-time factory fingerprint''}}). This calibration is achieved via the 
``\emph{Factory Fingerprint Builder}'' module, where a magnet is placed in each fine-grained cell in order and the fingerprint of the three-axes phone magnetometer readings are stored for each fine-grained cell. \textbf{\emph{This cumbersome process is only performed by us (\sys{}'s developers) once in a lifetime, and not by the applications developers or system user. In addition, this one-time calibration process is independent from future system's users, magnets, or cell phones that will be different from the ones used in this calibration process.}}

When the user downloads the \sys{}'s keyboard from the application store, and to compensate for the differences in the users' phone and magnet, a \textbf{\emph{simple one-minute}} calibration is performed, through the ``\textit{First Use Fingerprint Regenerator}'' module, to generate a device- and magnet-specific fingerprint. This is performed only during the first time \sys{} is used with a new phone or magnet.

During the actual system operation to detect keystrokes, the ``\textit{Keystroke Segmenter}'' segments keystrokes and then the ``\textit{Probabilistic Cell Estimator}'' uses the magnetometer's three-axes readings to estimate the magnet location over the \emph{fine-grained grid}, leveraging the stored fingerprint.  The estimated fine-grained cell is then mapped to the application-specific key using the layout stored in the \emph{Application-specific Key-mapping Database} through the ``\emph{Key Mapper}'' module. 

In all modes of operation, the \textit{Offset Compensator} module removes noise and environmental effects, which makes the system independent from the phone orientation and magnet polarity.

\begin{figure*}[!t]
\centering
        \begin{subfigure}[t]{0.29\textwidth}
                \centering
                \includegraphics[width=\textwidth]{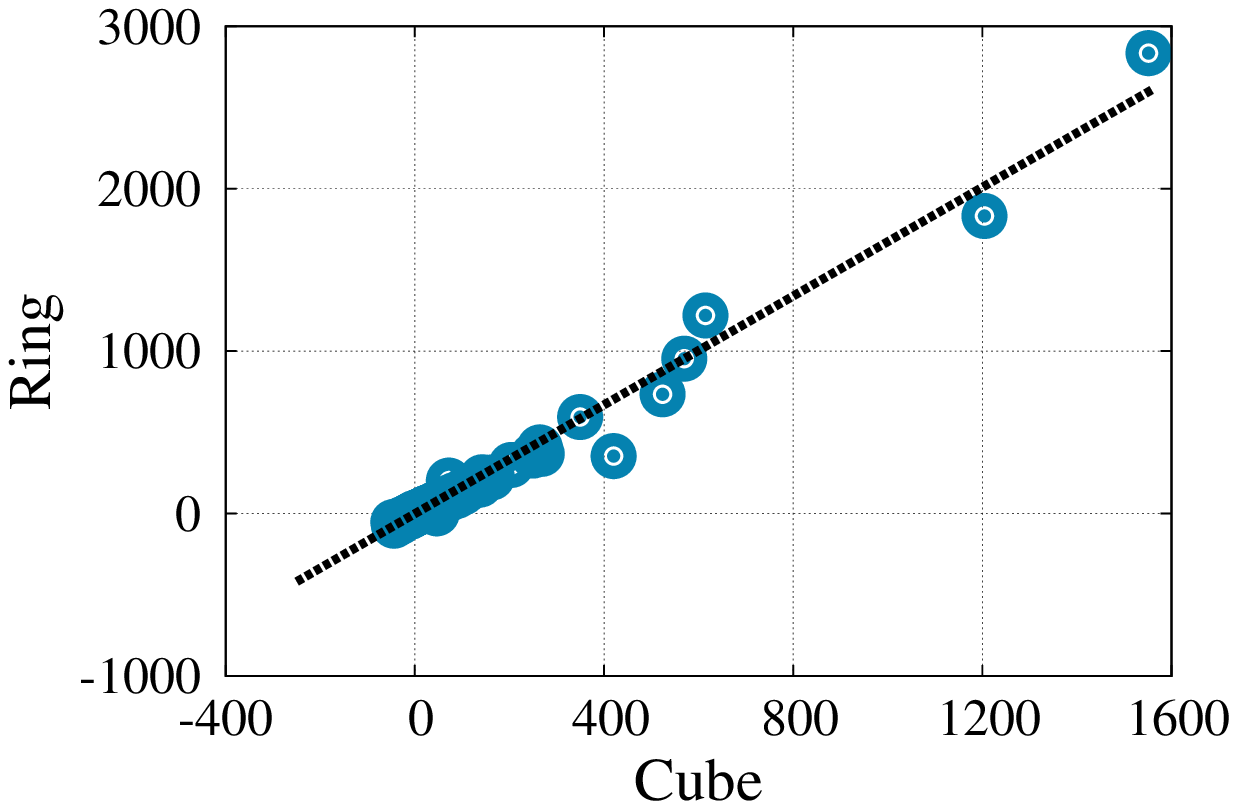}
                \caption{Magnetic field x-axis.}
                \label{fig:x-axis}
        \end{subfigure}~~
        \begin{subfigure}[t]{0.29\textwidth}
                \centering
                \includegraphics[width=\textwidth]{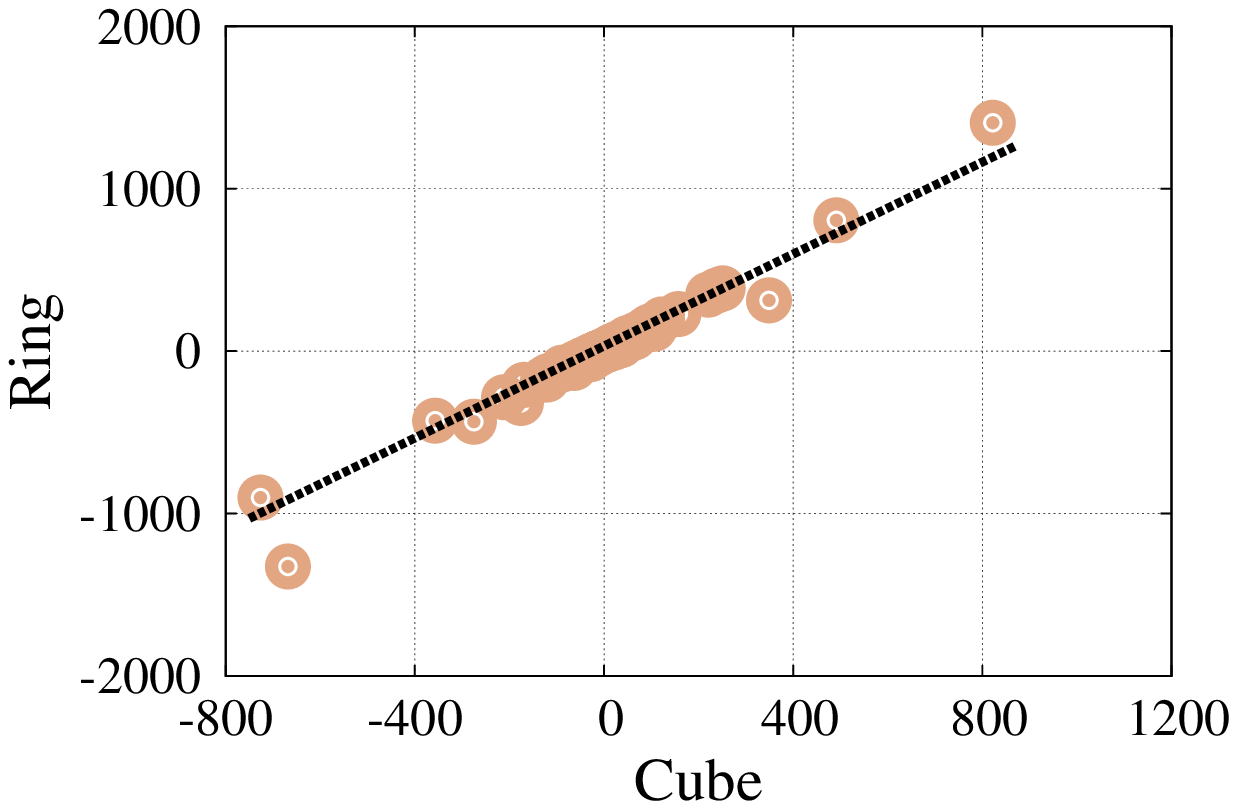}
                \caption{Magnetic field y-axis.}
                \label{fig:y-axis}
        \end{subfigure}~
        \begin{subfigure}[t]{0.29\textwidth}
                \centering
                \includegraphics[width=\textwidth]{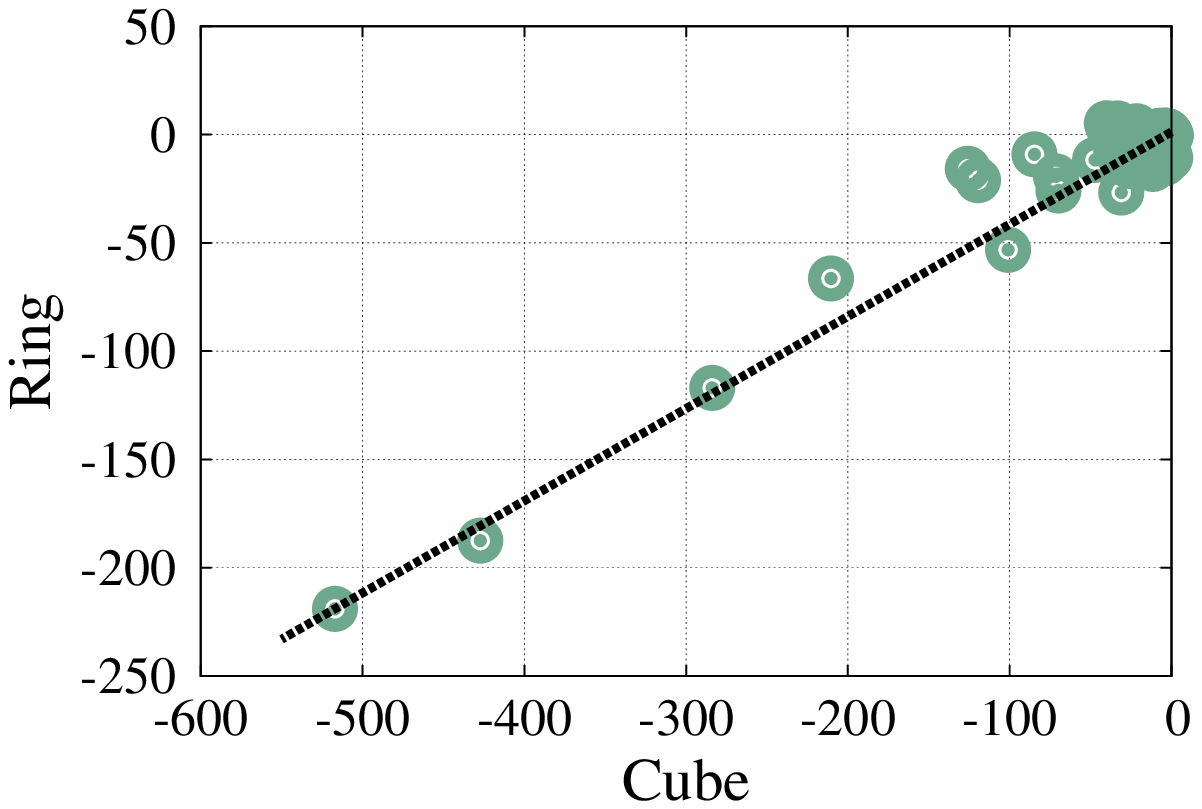}
                \caption{Magnetic field z-axis.}
                \label{fig:z-axis}
        \end{subfigure}

\caption{Mapping between the ring and cube magnets.}
\label{fig:ring_cube_mapping}
\vspace{-0.2in}
\end{figure*}

\begin{figure*}[!t]
\centering
        \begin{subfigure}[t]{0.29\textwidth}
                \centering
                \includegraphics[width=\textwidth]{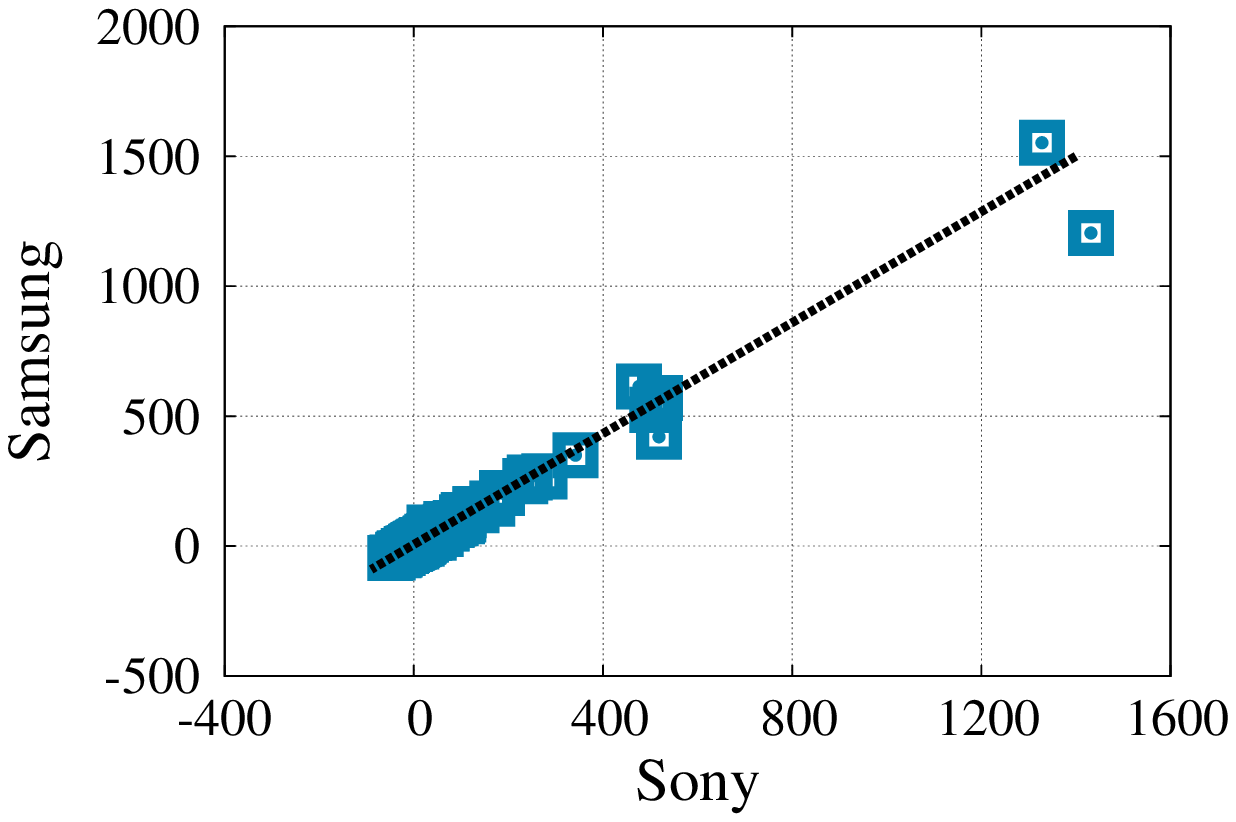}
                \caption{Magnetic field x-axis.}
                \label{fig:x-axis}
        \end{subfigure}~~
        \begin{subfigure}[t]{0.29\textwidth}
                \centering
                \includegraphics[width=\textwidth]{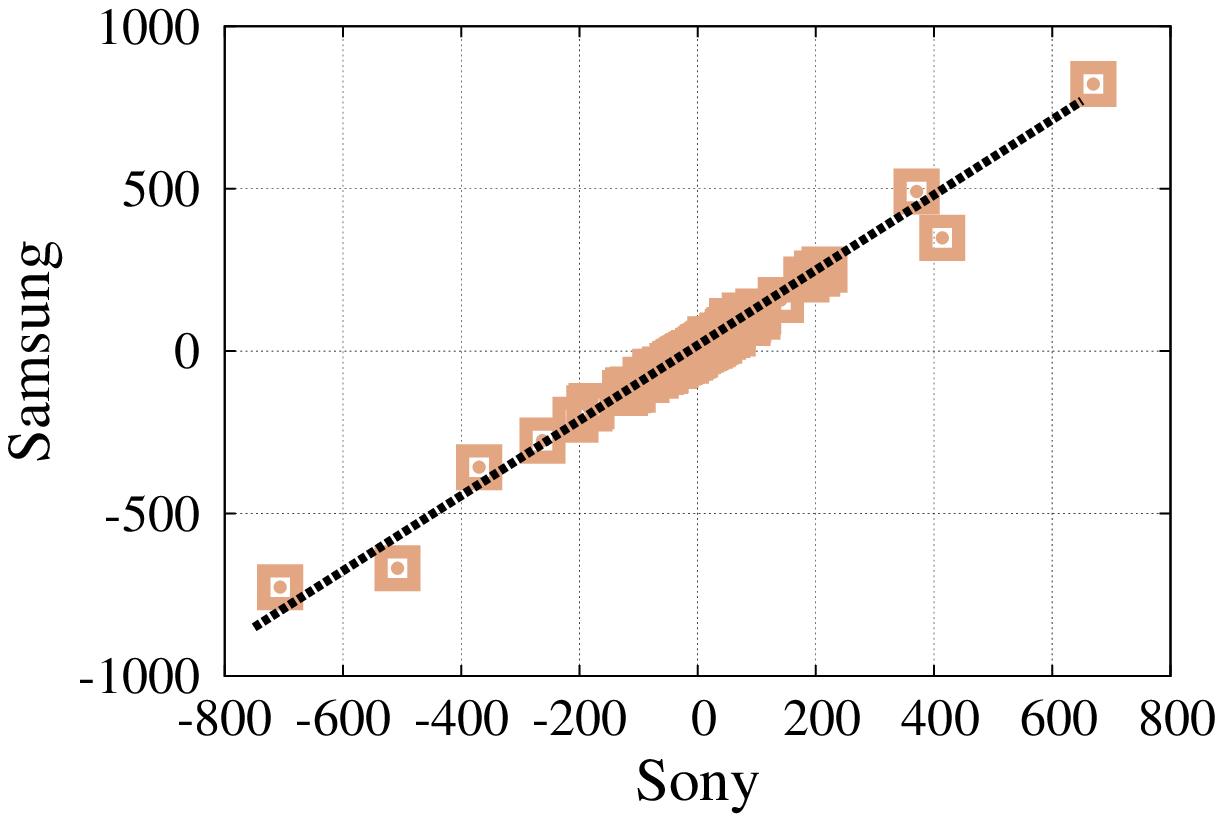}
                \caption{Magnetic field y-axis.}
                \label{fig:y-axis}
        \end{subfigure}~
        \begin{subfigure}[t]{0.29\textwidth}
                \centering
                \includegraphics[width=\textwidth]{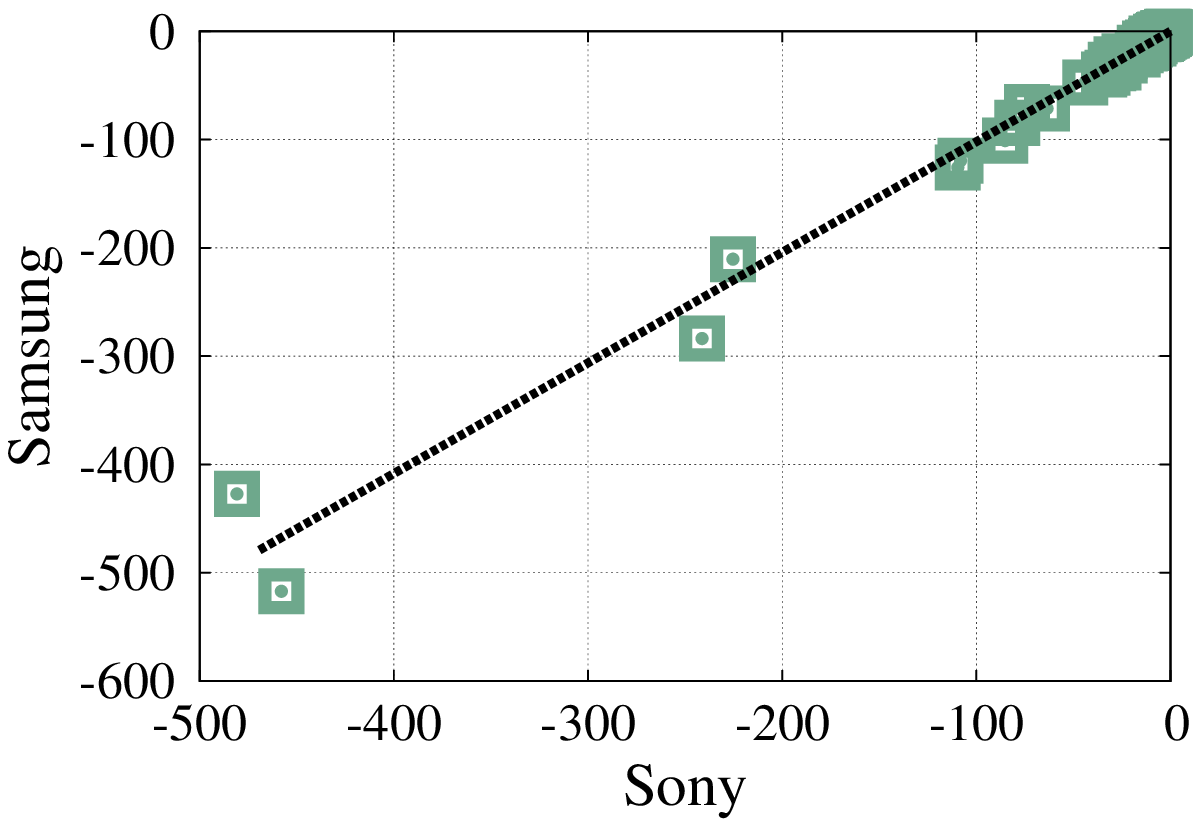}
                \caption{Magnetic field z-axis.}
                \label{fig:z-axis}
        \end{subfigure}

\caption{Mapping between Samsung and Sony phones}
\label{fig:samsung_sony_ring_mapping}
\vspace{-0.2in}
\end{figure*}

\subsection{Offline Calibration}

This section discusses the different calibration processes that are applied to make \sys{} independent from any surrounding environmental noise, changes in magnet polarity, and the characteristics of the used device or magnet. These are the responsibilities of the ``Offset Compensator'', ``Factory Fingerprint Builder'', ``First Use Fingerprint Regenerator'', and ``Polarity Handler'' modules.

\subsubsection{Offset Compensator}
\label{sec:offset_comp}

Since the Earth's magnetic field  ($B_{\textrm{Earth}}$) varies at different locations and the background magnetic field strength ($B_{\textrm{background}}$) dynamically changes, it is important for \sys{} to compensate for these changes to obtain a robust fingerprint. To do that, whenever the system is started, we remove the effect of these two varying sources of magnetic field by collecting samples without the existence of the \sys{}'s magnet for a few seconds (15 sec). Subtracting the average of this ``silence period'' from the magnetometer's readings leads to a fingerprint that is independent from the system location on Earth as well as the background magnetic interference. 
More formally, from Equation~\ref{eq:Bt_wonoise}, the measured magnetic field after silence removal becomes:

\begin{eqnarray}
B_P &=& R_x(\phi)R_y(\theta)Rz(\psi) [B_T- B_{\textrm{silence}}] \nonumber \\
    &=& R_x(\phi)R_y(\theta)Rz(\psi) B_{\textrm{magnet}}
\label{eq:Bt_offset}
\end{eqnarray}

where $B_{\textrm{silence}}= B_{\textrm{Earth}}+ B_{\textrm{background}}$.

Finally, we note that the phone location in \sys{} is fixed on the board relative to the grid (Figure~\ref{fig:fingerprint_map}), which 
makes \sys{} \textbf{\emph{independent of the phone orientation}} as the phone attitude is always fixed relative to the magnet. In other words, the phone, magnet, and the board move together as a rigid body when the phone is rotated in 3D.

\subsubsection{Factory-based Fingerprint Building}

\sys{} stores the magnet effect/signature on the embedded magnetometer in the phone, in different board cells relative to the phone. Specifically, the \textbf{once-in-lifetime} factory-based fingerprint $B_p$ is constructed by storing the histograms of the \emph{offset-free} magnetic field readings of each magnetometer axes from Equation~\ref{eq:Bt_offset}, and is stored for later retrieval by the other modules. 

Note that this fingerprint is constructed by us at the finest level ($2\textrm{cm}\times2\textrm{cm}$), so that its cells can be easily combined into higher-level customized keyboards by the applications' designers or users through the ``Keyboard Designer'' module.

\subsubsection{First Time Use Fingerprint Regeneration}

Since \sys{} can work with any phone or magnet that differ from the ones used in the factory calibration, the \textit{First Use Fingerprint Regenerator} module handles this \textbf{only once during the first time use}. In particular, when the user downloads \sys{} from the application store, a simple calibration is performed to map the factory-based fingerprint to a new one that is specific to the user's mobile and magnet.

Figure~\ref{fig:ring_cube_mapping} shows the relation between the magnetometer readings for two different magnets in shape and strength when positioned at the same location. The figure shows that there is a linear relation between the different magnets. 
Similarly, Figure~\ref{fig:samsung_sony_ring_mapping} shows that the relation between the magnetometer readings of two different phone sensors using the same magnet is also linear. 
To estimate the parameters of the linear mapping relation, \sys{} collects samples at specific cells in the grid. We find that collecting data from only two cells is sufficient for estimating the parameters accurately over duration of 15 sec for each cell.
\textbf{\emph{Therefore, the fingerprint regeneration process takes less than one minute and is required only once for any new device or magnet.}}

\subsubsection{Handling Polarity}

Since the user can wear or hold the magnet arbitrarily, there can be a change of polarity (sign) of the magnetometer readings as show in Figure~\ref{fig:changing_polarity_effect}. To handle this oscillation, \sys{} uses the readings at a \emph{single} known cell to determine if there is a change of polarity or not. This is usually achieved by clicking a known cell at the beginning, e.g. the power button of the calculator. 

\begin{figure}[!t]
\centering
\includegraphics[width=2.5in]{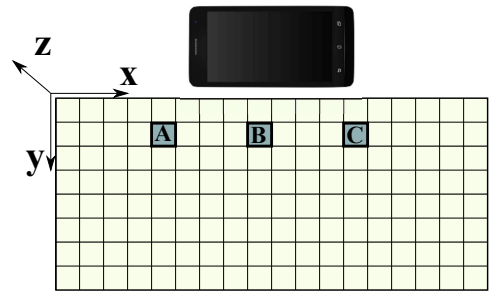}
\caption{Fine-grained fingerprint map - each spot has area of 2cm$\times$2cm. 
}
\label{fig:fingerprint_map}
\vspace{-0.2in}
\end{figure}
\begin{figure}[!t]
\centering
        \begin{subfigure}[t]{0.19\textwidth}
                \centering
                \includegraphics[width=\textwidth]{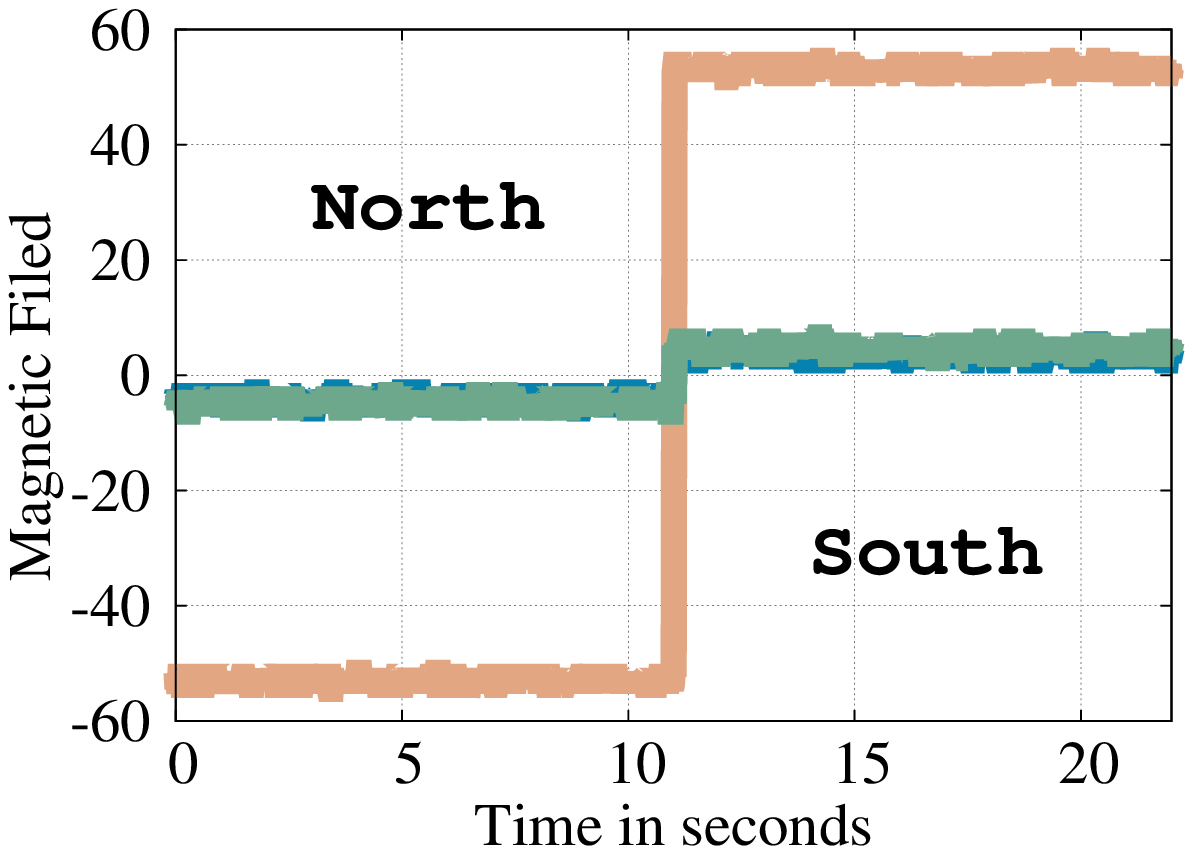}
                \caption{Spot A.}
                \label{fig:spot23}
        \end{subfigure}
        \begin{subfigure}[t]{0.19\textwidth}
                \centering
                \includegraphics[width=\textwidth]{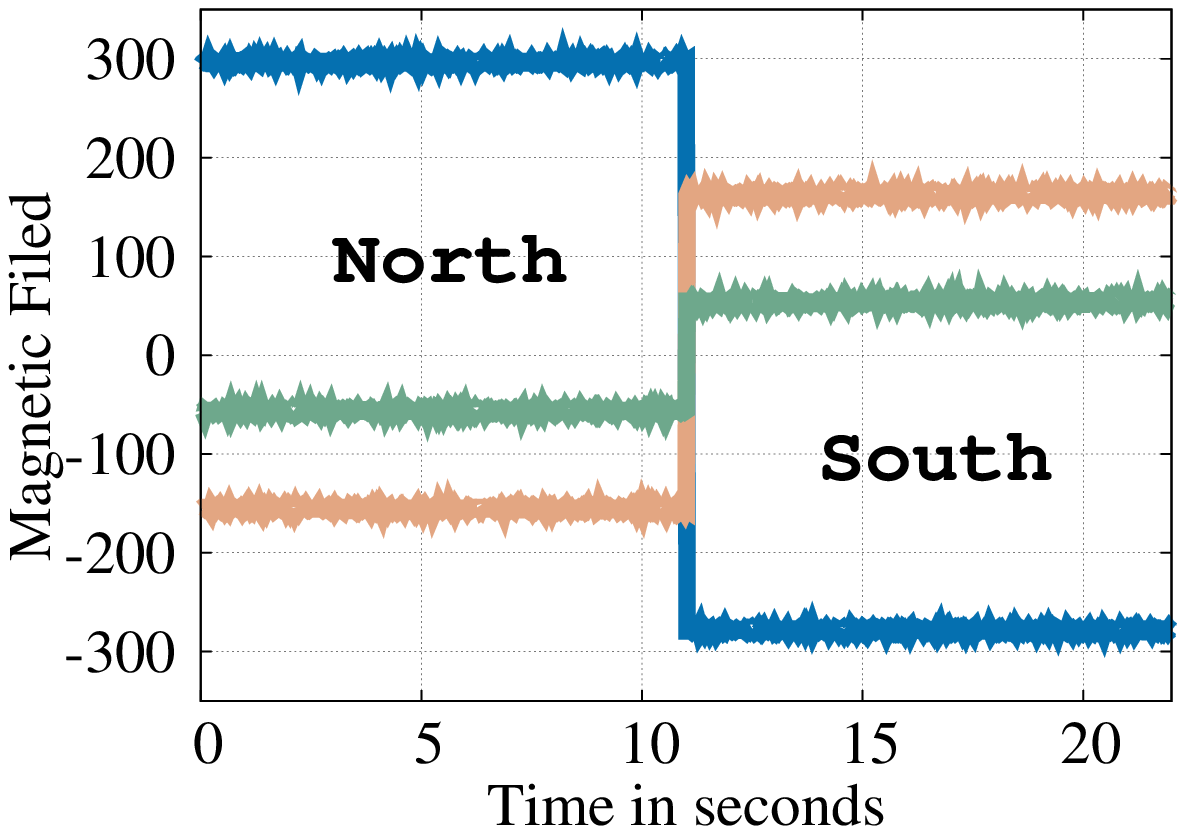}
                \caption{Spot B.}
                \label{fig:spot25}
        \end{subfigure}
\caption{Changing magnet polarity effect on the $A$ and $B$ spots from Fig.~\ref{fig:fingerprint_map}. The blue, red, and green lines represent the x, y, and z-axis readings respectively.}
\label{fig:changing_polarity_effect}
\vspace{-0.2in}
\end{figure}

\subsection{Keystroke Recognition}

This section discusses how the magnet strokes are translated to clicked application buttons.

\subsubsection{Keystroke Segmentation} 
\label{sec:motion_det}

\begin{figure}[!t]
\centering
\includegraphics[width=3.2in]{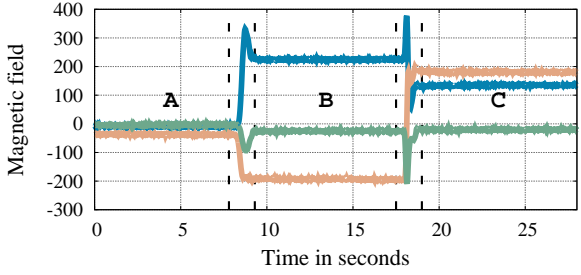}
\caption{Keystroke segmentations when the magnets click spots $A$, $B$, and then $C$ shown in Fig.~\ref{fig:fingerprint_map}. The transition periods are characterized by a high variance compared to when the key is stationary on a spot. Note that the required keystroke time is less than one second. } 
\label{fig:segmintation}
\vspace{-0.2in}
\end{figure}

Since \sys{} tracks the click of discrete keys, we have two states for the magnet: moving and stationary. The ``moving'' state occurs when the user is moving the magnet to click on a different key. The ``stationary'' state occurs when the magnet is stable over a button key. To increase the accuracy and robustness of estimation, \sys{} leverages the magnetic field readings variance to detect the state. The idea is that when the magnet is moving, there is a higher variance in the magnetic field strength as shown in Figure~\ref{fig:segmintation}. Therefore, a simple threshold-based approach can be used to accurately estimate when the magnet is stationary over the board as compared to moving between keys. 

\subsubsection{Probabilistic Cell Estimation}
We now discuss how \sys{} determines the fine-grained cell the magnet is placed over, based on the magnetometer readings and the regenerated fingerprint. 
Let $n$ represent the total number of streams in the system (typically the three-axes streams of the magnetometer sensor).  Given a received magnetic field strength vector $s = (s_1,...,s_n)$, we want to find the cell $x^* \in \mathbb{X}$ that maximizes the probability $P(x|s)$, where $\mathbb{X}$ is the set of all virtual grid cells. Mathematically, using Bayesian inversion, this can be written as:

\begin{equation}
x^* = \argmax_x P(x|s) = \argmax_x [\frac{P(s|x).P(x)}{P(s)}]
\label{equ:bayesian_inversion}
\end{equation}

$P(s)$ can be factored out as it is independent from $x$. Assuming all cells have the same probability, the $P(x)$ term can be factored out from the equation\footnote{If the distribution of $P(x)$ is known, it can be used in the equation.}. The equation then becomes:
\begin{equation}
x^* = \argmax_x P(x|s) = \argmax_x [P(s|x)]
\end{equation}

$P(s|x)$ can be calculated from the calibrated fingerprint histogram of each cell $x$ as:
\begin{equation}
P(s|x) = \Pi^n_{i=1} P(s_{i}|x)
\end{equation}

The last equation considers only one sample from each stream. However, a number of samples can be used in a single estimate to further improve the performance. If we have $m$ successive samples, $P(s|x)$ can be expressed as follows:
\begin{equation}
P(s|x) = \Pi^n_{i=1}\Pi^m_{j=1} P(s_{i,j}|x)
\label{equ:sample_stream}
\end{equation}

where $s_{i,j}$ represents the $j^{th}$ sample from the $i^{th}$ stream.

Thus, given the magnetic field strength vector $s$, the previous equations return the cell $x$ that has the maximum probability.

To further improve the localization accuracy, especially for coarser-grained keyboards when keys are grouped, \sys{} uses the spatial averaging technique to obtain a location estimate in the continuous space. \textbf{Spatial averaging} fuses the most probable $k$ cell centroids rather than the single top location. Each location of the $k$ locations is weighted with its probability in the final estimation. So the estimated location $x$ would be represented as:

\begin{equation}
x = \frac{\sum^{k}_{i=1} P(i).x_i}{\sum^{k}_{i=1} P(i)}
\end{equation}

Note that the estimated location $x$ need not to be one of the fingerprint map locations.

\subsubsection{Application Grid Mapping - Keyboard Designer}
Independently, the application developer or the user can map between the fine-grained fingerprint and the application board using a simple GUI module. One button in the application grid can be mapped to several spots in the fine-grained fingerprint as shown in Figure~\ref{fig:app_grid_mapping}. Also different buttons can be designed with different sizes, e.g. button `=' covers $8$ fine-grained spots while button `0' covers $4$ fine-grained spots. Such specifications guarantee the flexibility of the keyboard design process.

\begin{figure}[!t]
\centering
        \begin{subfigure}[t]{0.24\textwidth}
                \centering
                \includegraphics[width=\textwidth]{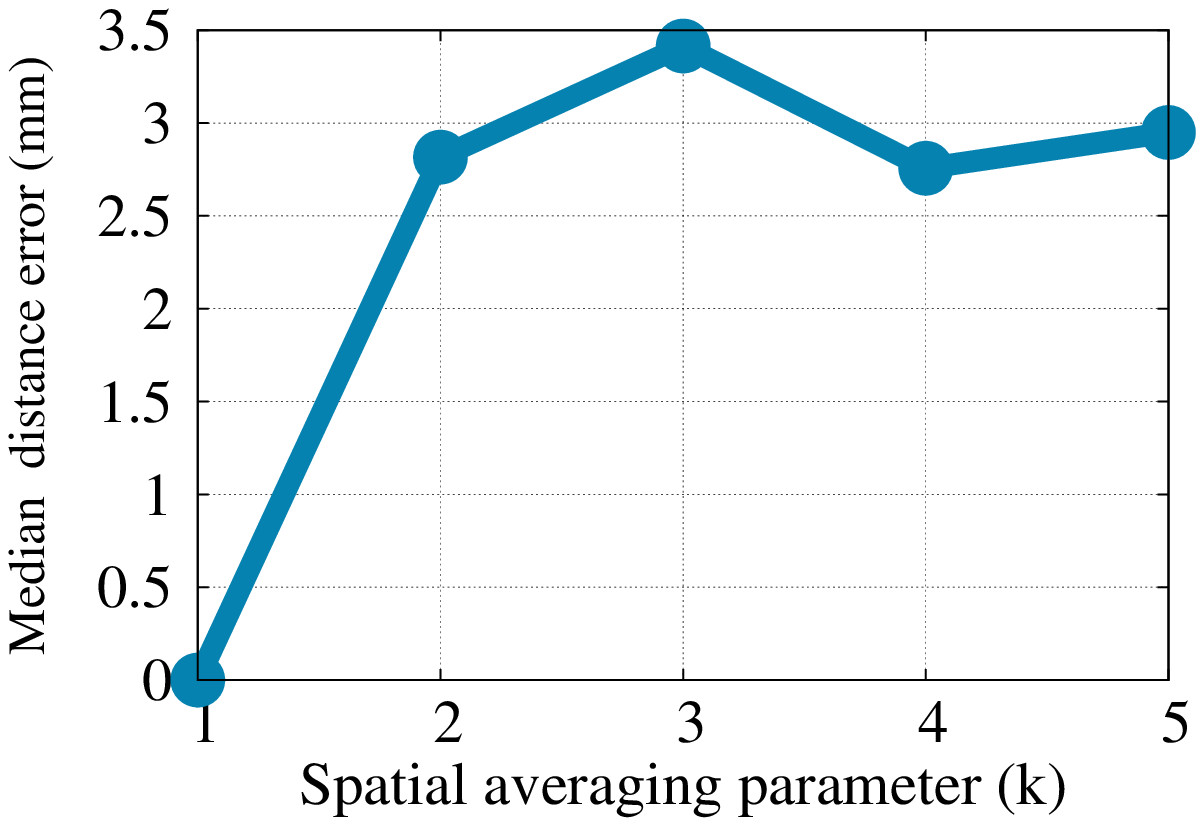}
                \caption{Discrete.}
                \label{fig:k_discrete}
        \end{subfigure}~~
        \begin{subfigure}[t]{0.24\textwidth}
                \centering
                \includegraphics[width=\textwidth]{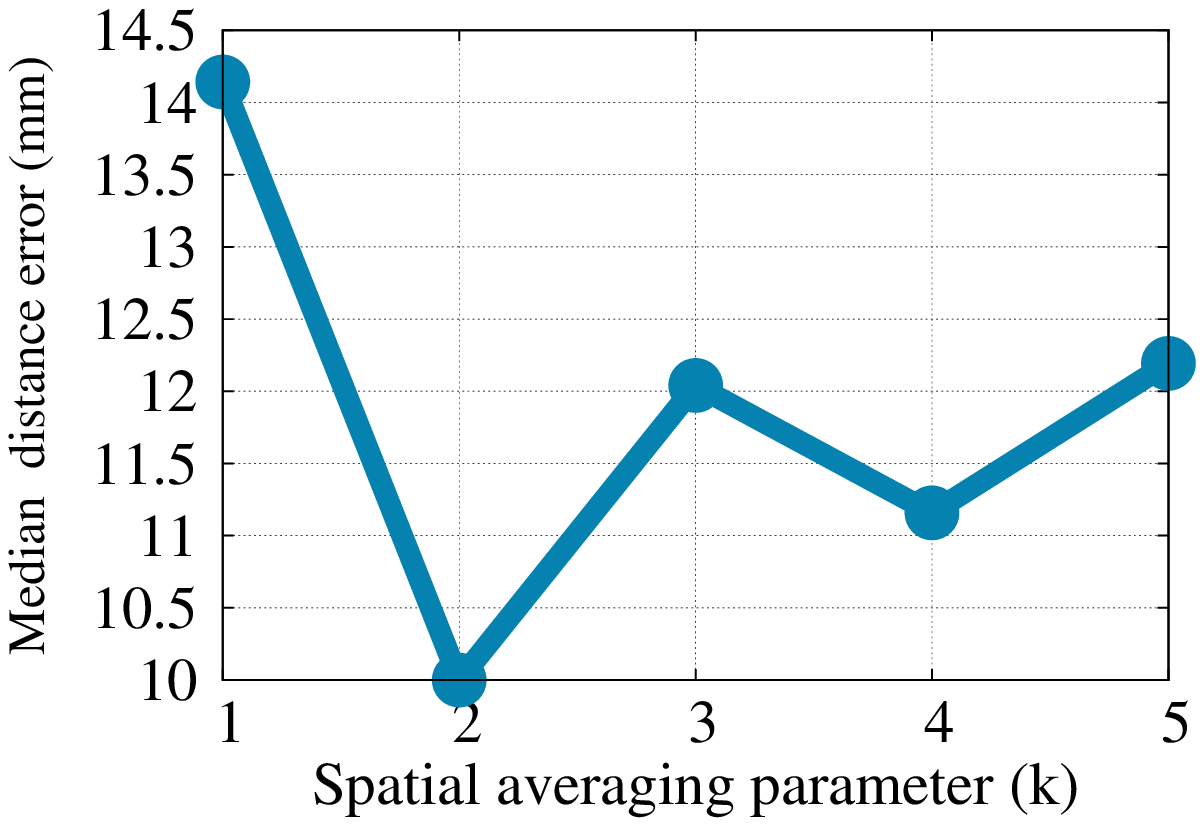}
                \caption{Continuous.}
                \label{fig:k_continuous}
        \end{subfigure}~
\caption{Effect of the spatial averaging parameter ($k$).}
\label{fig:spatial_averaging_parameter}
\vspace{-0.2in}
\end{figure}

\begin{figure*}[!t]
\minipage[t]{0.33\textwidth}
 \includegraphics[width=\linewidth]{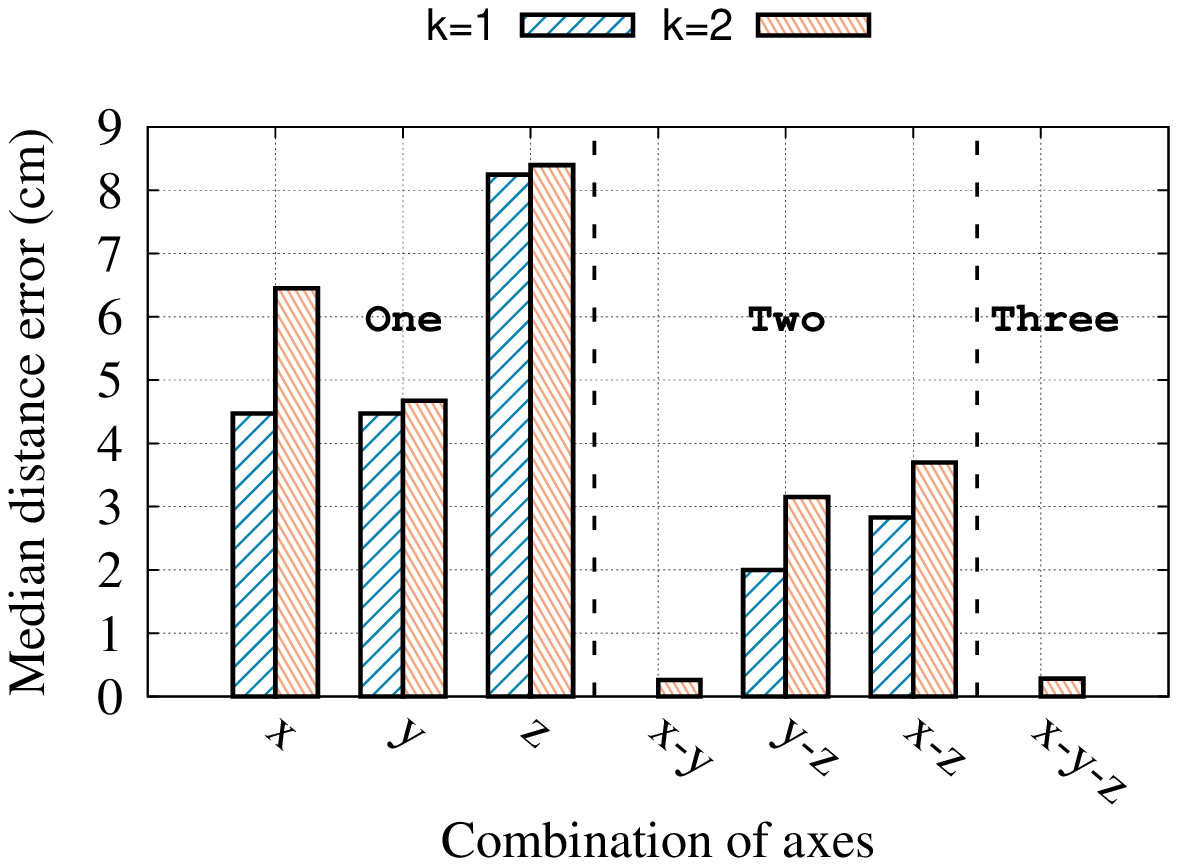}
 \caption{Effect of different combinations of magnetometer axes (x, y, z) on magnet location estimation accuracy.}
 \label{fig:axes_combinattions}
\endminipage\hfill~
\minipage[t]{0.32\textwidth}
  \includegraphics[width=\linewidth]{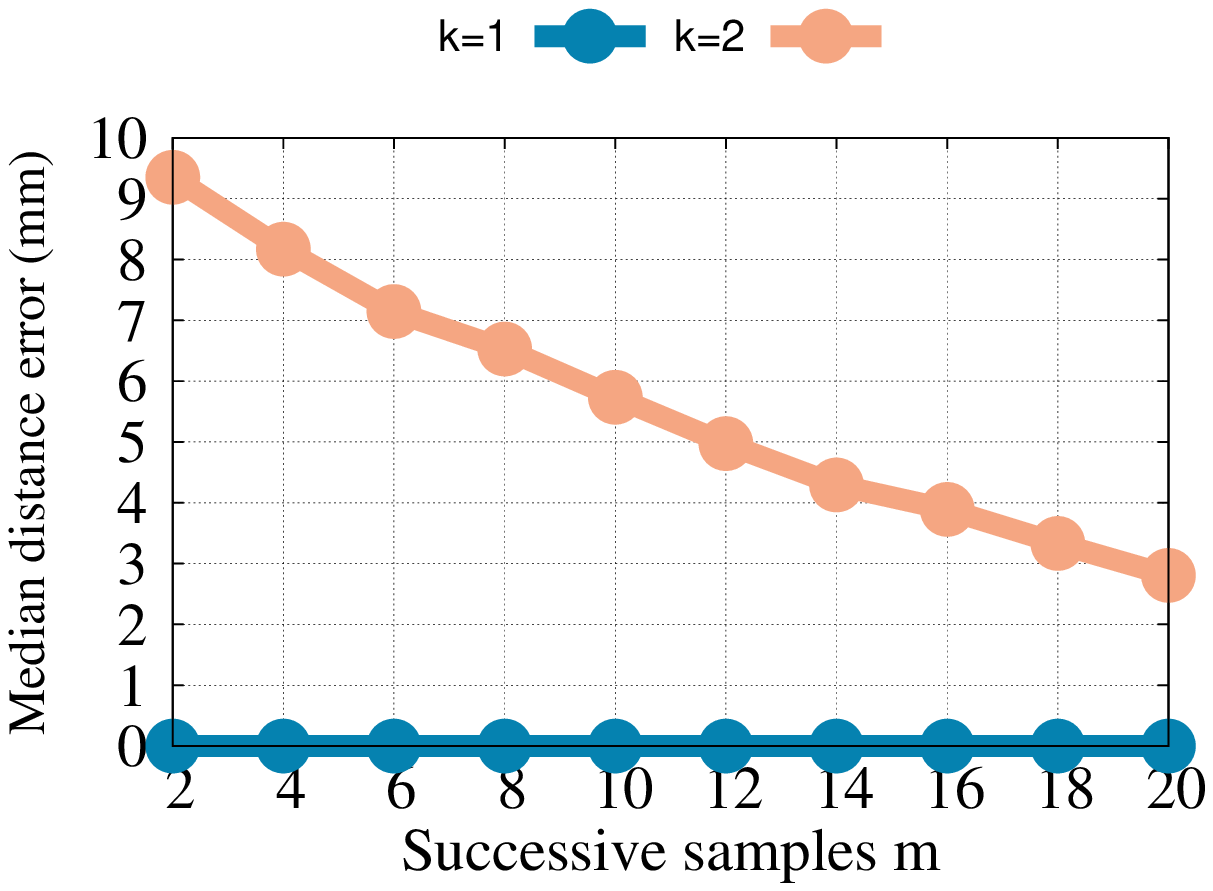}
 \caption{Effect of successive input samples ($m$).} 
 \label{fig:successive_samples}
\endminipage\hfill~
\minipage[t]{0.32\textwidth}%
\includegraphics[width=\linewidth]{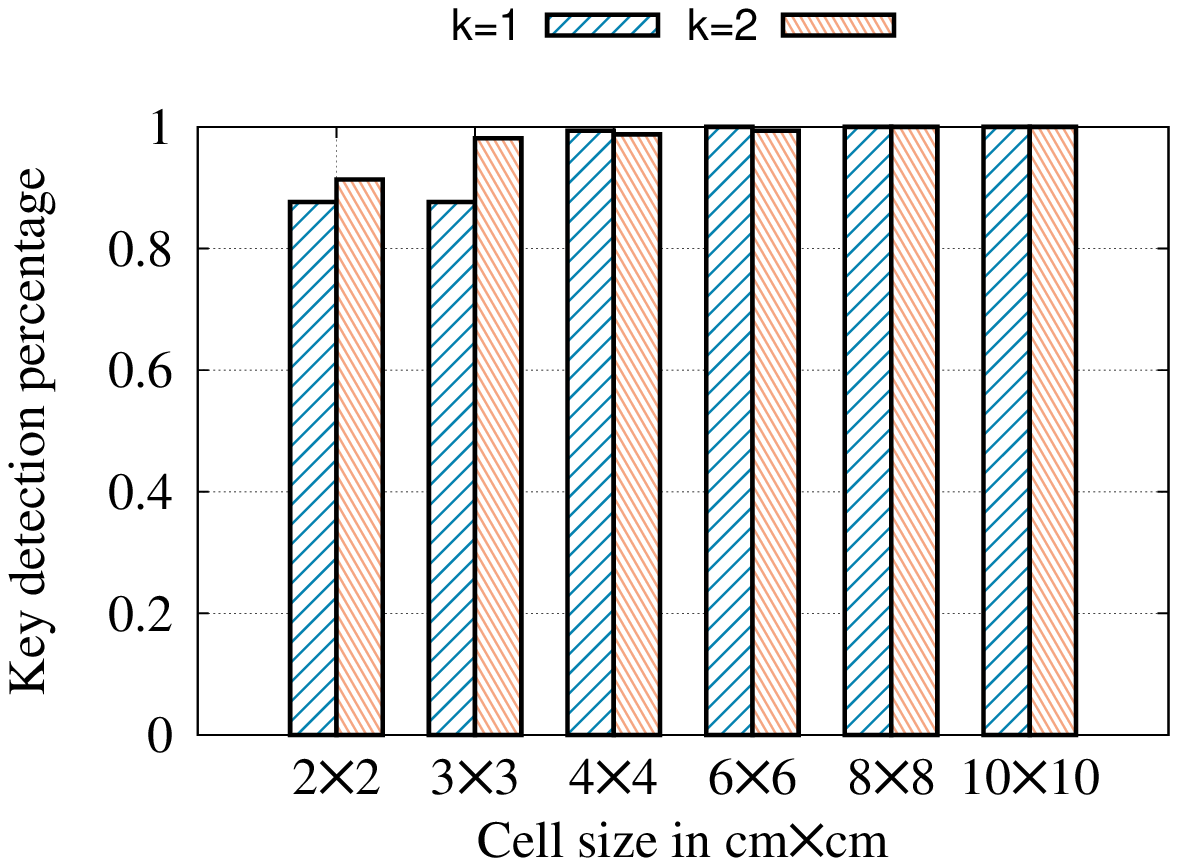}
\caption{Cell size (density) effect on system performance.}
\label{fig:density}
\endminipage
\vspace{-0.1in}
\end{figure*}

\section{Evaluation}
\label{sec:evaluation}

In this section, we study the performance of \sys{}. We start by describing the experimental setup, followed by studying the effect of different parameters on the key detection accuracy. We then study the ability of \sys{} to adapt to different magnets and mobile phones. Finally, we evaluate the performance of \sys{} in a typical application scenario.

\subsection{Experimental Setup}
We extensively evaluate \sys{} in different environments, with different users, over different days and times, and with different surrounding humans activities. We experiment with three Android phones: a Sony Z2, a Samsung S4, and a Nexus 4. We utilized different magnets with different shapes and strengths as shown in Table ~\ref{tab:diff_magnets_loc_erro}: a ring magnet (865 tesla), a cube magnet (971 tesla), and a magnet attached to a toy (114 tesla). The magnetometer sampling rate is $50$ Hz in all experiments (default for the Android API). 

To construct the fingerprint, we use a board of size 36cm$\times$16cm as shown in Figure~\ref{fig:fingerprint_map}. The board is divided into cells of size 2cm$\times$2cm each, leading to a total of 144 cells. We experimented with different boards \textbf{orientations} and found no effect on system accuracy due to the rigid body movement (Section~\ref{sec:offset_comp}). 
Additionally, we conduct experiments with two different magnet orientations, north and south, relative to the mobile phone. For each cell, the samples are collected for 15 sec (sufficient samples to construct stable histograms). 
Table~\ref{tab:def_parameters} summarizes the different parameters used in our experiments and their nominal values.

\begin{table}[!t]
    \centering
    \begin{tabular}{|l|c|c|} \hline
      \textbf{Parameter} & \textbf{Range} & \textbf{Nominal Value} \\ \hline \hline
      Streams ($n$) &  \specialcell{$x$, $y$, $z$,  $x$-$y$,  $x$-$z$, \\$y$-$z$, $x$-$y$-$z$}  & $x$-$y$-$z$ \\ \hline
      Succ. samples ($m$) & $2$ - $20$ & $20$ $\textrm{sample}$ \\ \hline
      Window  size ($w$) & $5$ - $30$ & $5$ $\textrm{sample}$ \\ \hline
     Spatial avg. ($k$) & $1$ - $5$ & $1, 2$ \\ \hline
     Magnets & \specialcell{Cube (971 tesla), \\ Ring (865 tesla),\\ Toy (114 tesla)} & Ring \\ \hline
     Phones & Sony Z2, Samsung S4, Nexus 4 & Sony Z2 \\ \hline
  \end{tabular}
    \caption{Parameter default values.}
    \label{tab:def_parameters}
    \vspace{-0.2in}
\end{table}

\subsection{Effect of Different Parameters} 
This section quantifies the impact of the different parameters on the system cell estimation error. The parameters include spatial averaging parameter ($k$), number of data streams ($n$), number of successive input samples ($m$)
, and virtual grid density ($d$). 
The subsection ends by quantifying the overall key detection accuracy. 

\subsubsection{Impact of the spatial averaging parameter ($k$)}
\label{sec:impact_of_spatial_averaging_parameter_k}
Figure~\ref{fig:spatial_averaging_parameter} shows the effect of increasing parameter $k$ on the localization median distance error. For the discrete case (i.e. the test cases locations are one of the fingerprint cells centroids), the best accuracy occurs at $k=1$. This is intuitive as the effect of increasing $k$ is to smooth the output locations, which is only better in the continuous case (i.e. the test locations are located anywhere, including the cells boundary), which is indicated in the continuous case figure (subfigure b). The figure shows the best accuracy for continuous case is when $k=2$. The continuous case is important when the button size covers more than one cell. For the rest of this subsection we show the effect of different parameters when $k=1$ and $k=2$. 

\subsubsection{Impact of the number of used sensors streams ($n$)}
Figure~\ref{fig:axes_combinattions} shows the effect of using different combinations of the magnetometer axes on the median error of the magnet location. The figure shows that increasing the number of input streams leads to better accuracy. In addition, using the x-y streams only approximately has the same 
median distance error as using the three x-y-z streams. This is intuitive as the magnet is moved in x-y plane, so the $z$ dimension does not provide relevant information for magnet localization. This can be used to reduce the computation power of \sys{}; 
but not the sensor consumed power as the three streams are provided all the time and cannot be disabled individually. 
In addition, if multiple mobile devices are available, they can be leveraged to further enhance accuracy.

The $z$ stream, on the other hand, is important for detecting the magnet motion as we quantify in Section~\ref{sec:motion_res}.

\begin{figure}[!t]
\centering
\includegraphics[width=3.5in]{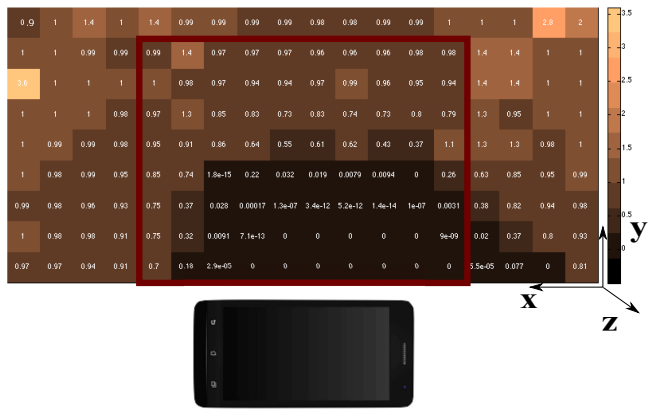}
\caption{Heatmap of the \sys{} keyboard error. The dark color denotes high accuracy. 
The spots near the magnetometer sensor have better accuracies (due to the higher magnet strength). 
 The red rectangle represents the covered area by the calculator case study in Section~\ref{sec:case_study}.}
\label{fig:heatmap}
\vspace{-0.2in}
\end{figure}

\subsubsection{Impact of successive input samples ($m$)}

Figure~\ref{fig:successive_samples} shows the effect of increasing the number of input samples used from each stream per keystroke estimate on the system accuracy. The figure shows that the accuracy increases with $m$. However, as $m$ increases, the latency required per key estimate of the system increases linearly. Therefore, a balance is required between the accuracy and latency of the system. We use a default value of $20$ for parameter $m$ to balance these two effects. 

\begin{figure*}[!t]
\minipage{0.32\textwidth}
 \includegraphics[width=\linewidth]{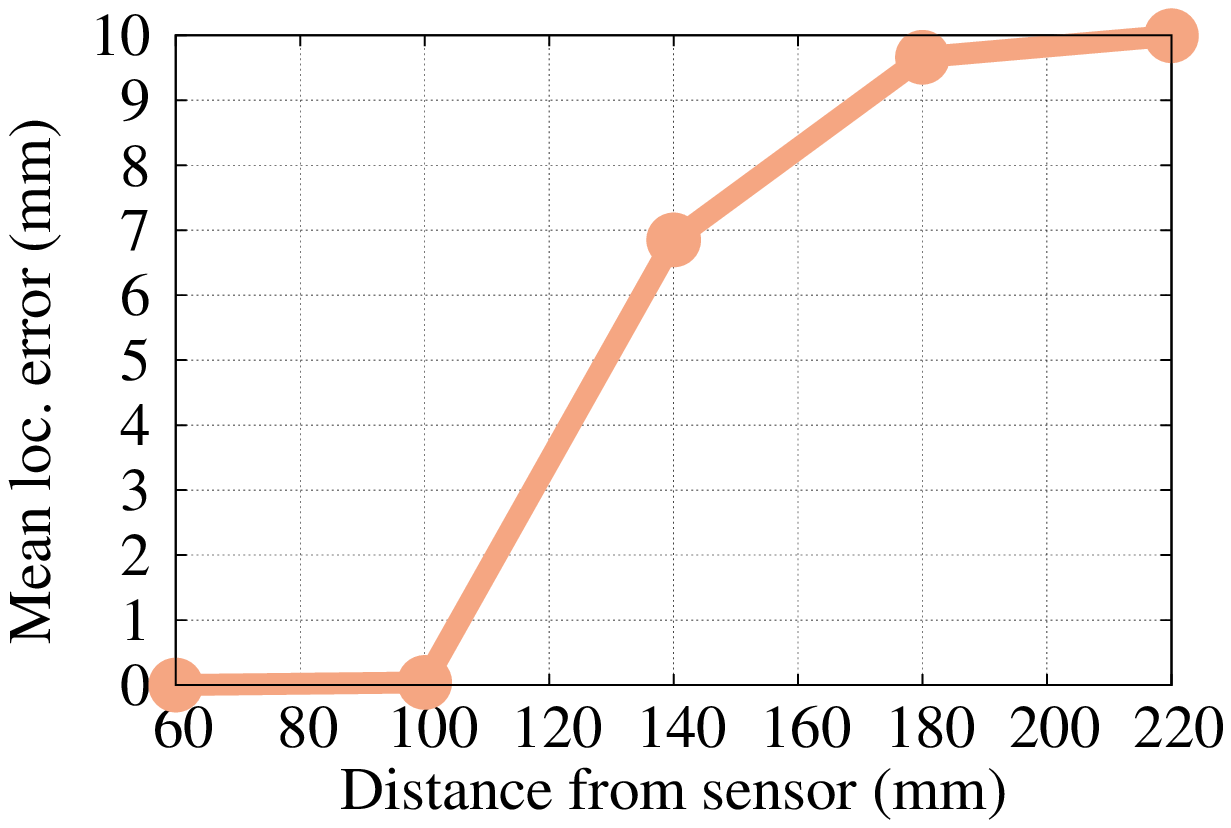}
\caption{
Accuracy as a function of the distance between the magnet and phone sensor.}
\label{fig:distance_accuracy}
\endminipage\hfill~
\minipage{0.32\textwidth}
  \includegraphics[width=\linewidth]{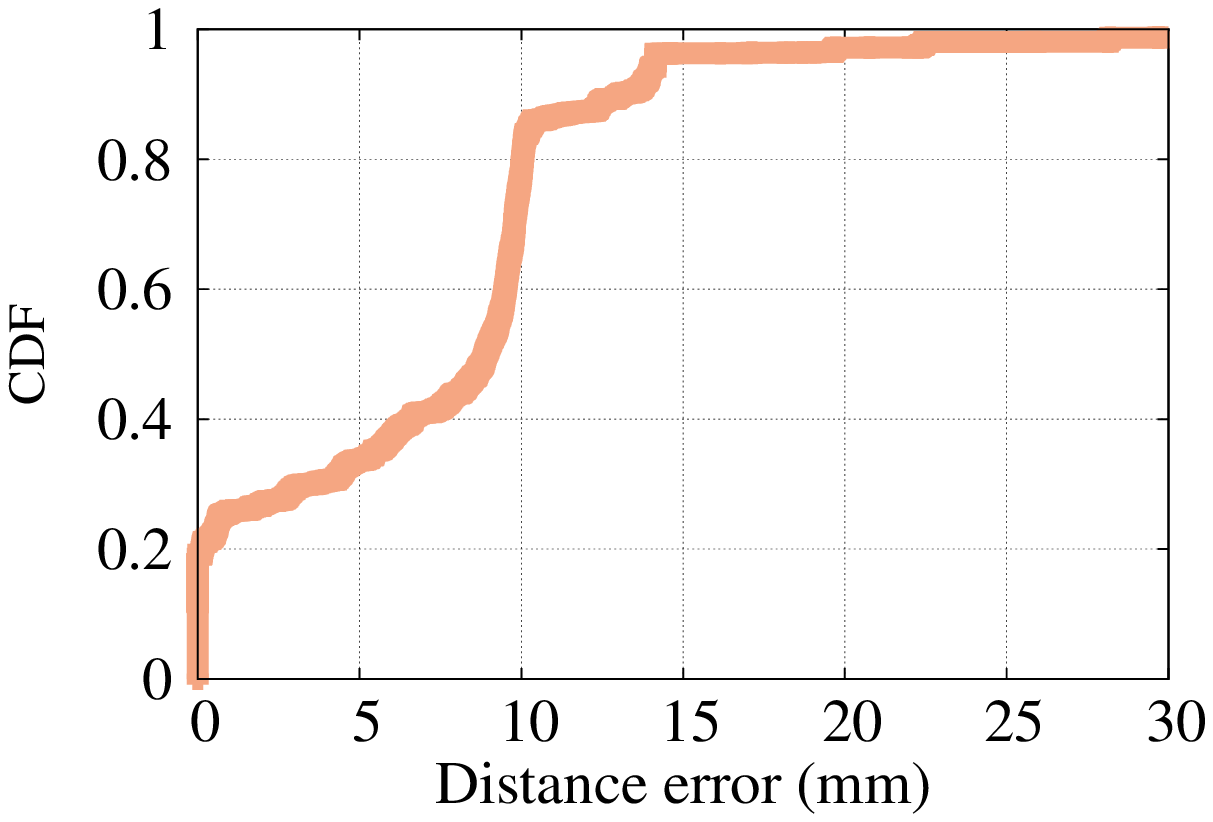}
\caption{CDF of magnet localization distance error.}
\label{fig:cdf}
\endminipage\hfill~
\minipage{0.32\textwidth}%
 \includegraphics[width=\linewidth]{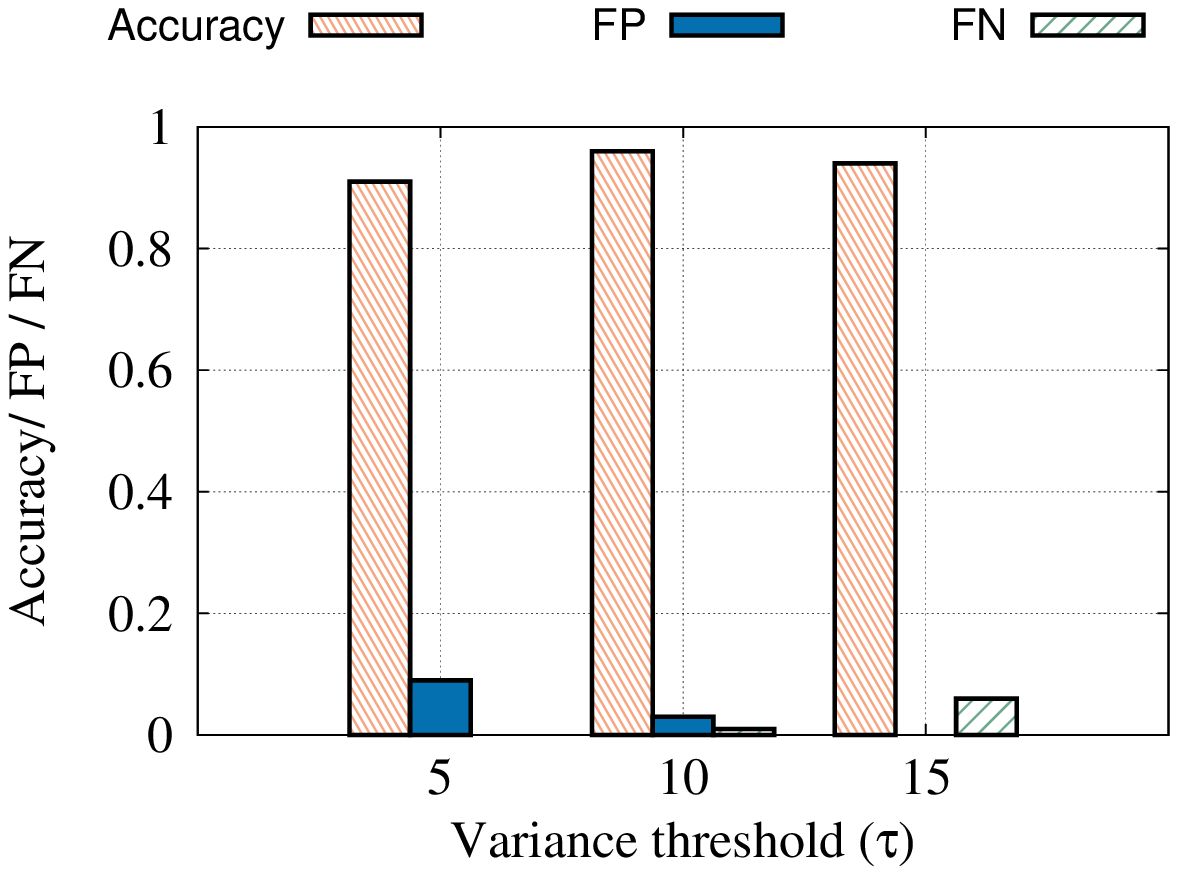}
\caption{Key segmentation accuracy as a function of the variance threshold ($\tau$).}
\label{fig:motion_detection_accuracy}
\endminipage
\vspace{-0.2in}
\end{figure*}

\subsubsection{Impact of the virtual grid density ($d$)}
Figure~\ref{fig:density} shows the impact of changing the virtual grid size (density) on the key detection accuracy. As expected, the higher the grid density the lower the detection accuracy due the more similar signature between adjacent dense cells. Nevertheless, \sys{} can achieve more than 91\% detection accuracy even with the highest cell density of one cell every 2cm$\times$2cm. In addition, increasing the $k$ parameter leads to further enhancement in accuracy (as discussed in Section~\ref{sec:impact_of_spatial_averaging_parameter_k}). Note the different metrics used in  Figure~\ref{fig:spatial_averaging_parameter} and Figure~\ref{fig:density}, which highlight the reason behind the contradiction between both figures in the 2cm$\times$2cm cell size case (considering that the 2cm$\times$2cm cell size is a discrete case). 

\subsubsection{Overall key detection accuracy}
Figure~\ref{fig:heatmap} shows the heat map of the detection error as a function of the key location in the virtual grid. The figure shows that the cells with the highest accuracies are those that are near to the magnetometer sensor. This is due to the stronger measured magnetic field near the phone sensor, which leads to better signal-to-noise ratio (Figure~\ref{fig:distance_accuracy}).

Figure~\ref{fig:cdf} further shows the CDF of mean distance error over the entire virtual grid. The figure shows that \sys{} can achieve 
less than 10mm median error. 
This highlights its applicability to a wide range of scenarios. 

\subsection{Key Segmentation: Motion State Detection}
\label{sec:motion_res}
Figure~\ref{fig:motion_detection_accuracy} shows the effect of the variance threshold $\tau$ described in Section~\ref{sec:motion_det} on the motion state detection accuracy.  The figure shows that the false positive rate decreases with the increase of the variance threshold while the false negative rate increases. The system has the best accuracy (defined as TP / (FP+FN+TP)) with a threshold of $10$.

\subsection{Effect of Using Different Magnets/Mobiles}
In this section, we evaluate the ability of \sys{} to generalize its fingerprint to different heterogeneous devices and magnets.

\subsubsection{Magnets heterogeneity}
To evaluate the ability of \sys{} to adapt to different magnets, we use three magnets with different shapes and strengths: Cube, Ring, and Car with strengths $971$, $865$, and $114$ tesla respectively. 
Table~\ref{tab:diff_magnets_loc_erro} summarizes the accuracy for all combinations of pairs of magnets used for training and testing. 

From the table, the ring and cube magnets fingerprints lead to the best accuracy. This is due to their high strength. The ring magnet has a slightly better performance as the cube magnet had a slightly irregular shape (was not a perfect cube). 
 Therefore, it is used as the ``master'' factory-based fingerprint of \sys{}. 

\subsubsection{Devices heterogeneity}
To evaluate the robustness to different mobile phones, we use three mobiles:  Sony Z2, Samsung S4, and Nexus 4. Table~\ref{tab:diff_mobiles_loc_erro} shows the accuracy. The table shows that by using the Sony Z2 phone as the master phone, \sys{} can achieve the best accuracy of less than 2.15 cm worst case mean error, highlighting its robustness to phone changes. 

\begin{table}[!t]
    \centering
    \begin{tabular}{|p{1.5cm}||c||c|c|c|} \hline
    \centering
      \textbf{\backslashbox{\scriptsize Test}{\scriptsize Train}}
      &\textbf{Metrics}&
     \begin{minipage}{.07\textwidth}
     \centering
     \includegraphics[width=8mm]{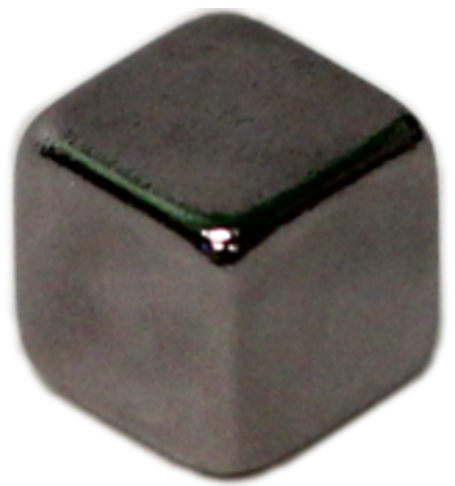}
     \end{minipage} &
     \begin{minipage}{.07\textwidth}
     \centering
     \includegraphics[width=8mm]{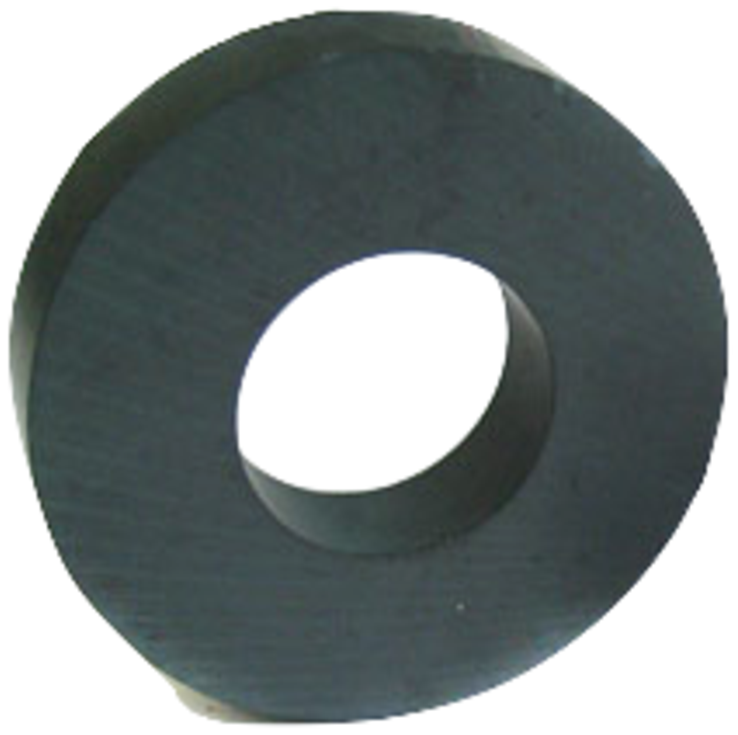}
     \end{minipage} &
     \begin{minipage}{.07\textwidth}
     \centering
      \includegraphics[width=8mm]{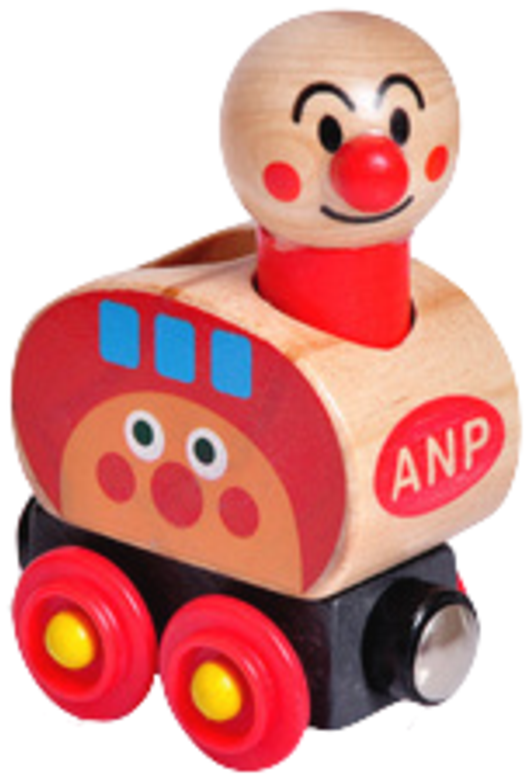}
      \end{minipage}  \\ \hline \hline

	\centering
     \multirow{3}{*}{ 
      \begin{minipage}{.15\textwidth}
      \includegraphics[width=8mm]{NIBCUB}
      \end{minipage}}
     &\textbf{Avg}& $0.13$ & $1.85$ & $2.38$\\
     &\textbf{50\%}& $0$ & $0$ & $2$\\
     &\textbf{75\%}& $0$ & $2.83$ & $4.47$  \\ \hline

      \multirow{3}{*}{
      \begin{minipage}{.15\textwidth}
      \includegraphics[width=8mm]{ring}
      \end{minipage}}
      &\textbf{Avg}&$2.06$ & $0.27$ & $2$\\
      &\textbf{50\%}&$0$ & $0$ & $0$\\
      &\textbf{75\%}&$2.83$ & $0$ & $4$\\ \hline 

      \multirow{3}{*}{
      \begin{minipage}{.15\textwidth}
      \includegraphics[width=8mm]{toy}
      \end{minipage}}
      &\textbf{Avg}&$2.17$ & $1.66$ & $0.87$\\
      &\textbf{50\%}& $2$ & $0$ & $0$\\
      &\textbf{75\%}&$4$ & $ 2.83$ & $1.41$\\ \hline
  \end{tabular}
\caption{Localization error for different pairs of magnets for training and testing. 
 Each cell contains three values: the mean, median, and 75th percentile of localization error.}    \label{tab:diff_magnets_loc_erro}
\end{table}

\begin{table}[!t]
    \centering
    \begin{tabular}{|p{1.5cm}||c|c|c|c|} \hline
      \textbf{\backslashbox{\scriptsize Test}{\scriptsize Train}}
      &\textbf{Metrics}&
      \begin{minipage}{.07\textwidth}
      \centering
      \includegraphics[height=8mm]{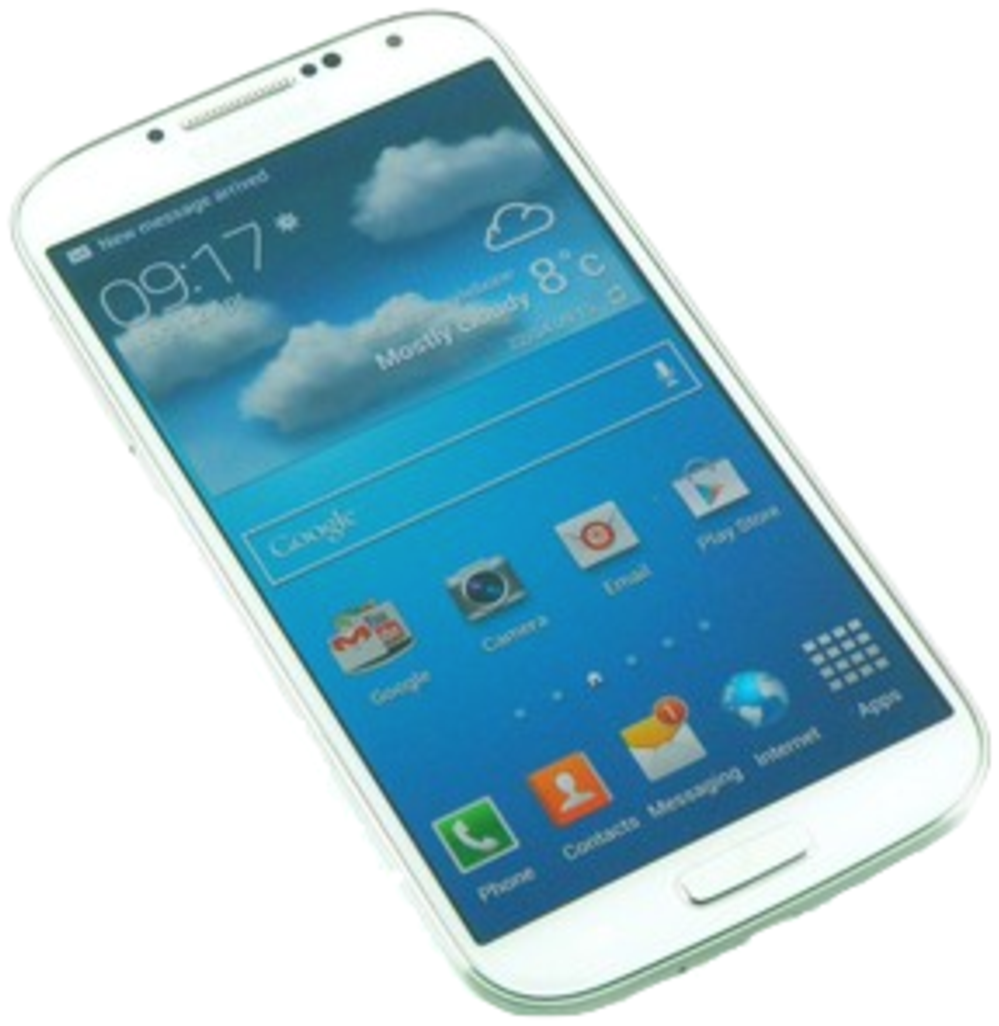}
      \\Sams. S4
      \end{minipage} &
      \begin{minipage}{.07\textwidth}
      \centering
      \includegraphics[height=8mm]{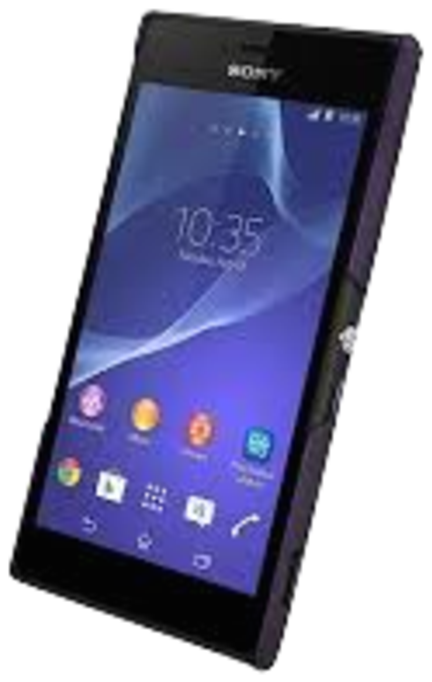}
      \\Sony Z2
      \end{minipage}&
      \begin{minipage}{.07\textwidth}
      \centering
      \includegraphics[height=8mm]{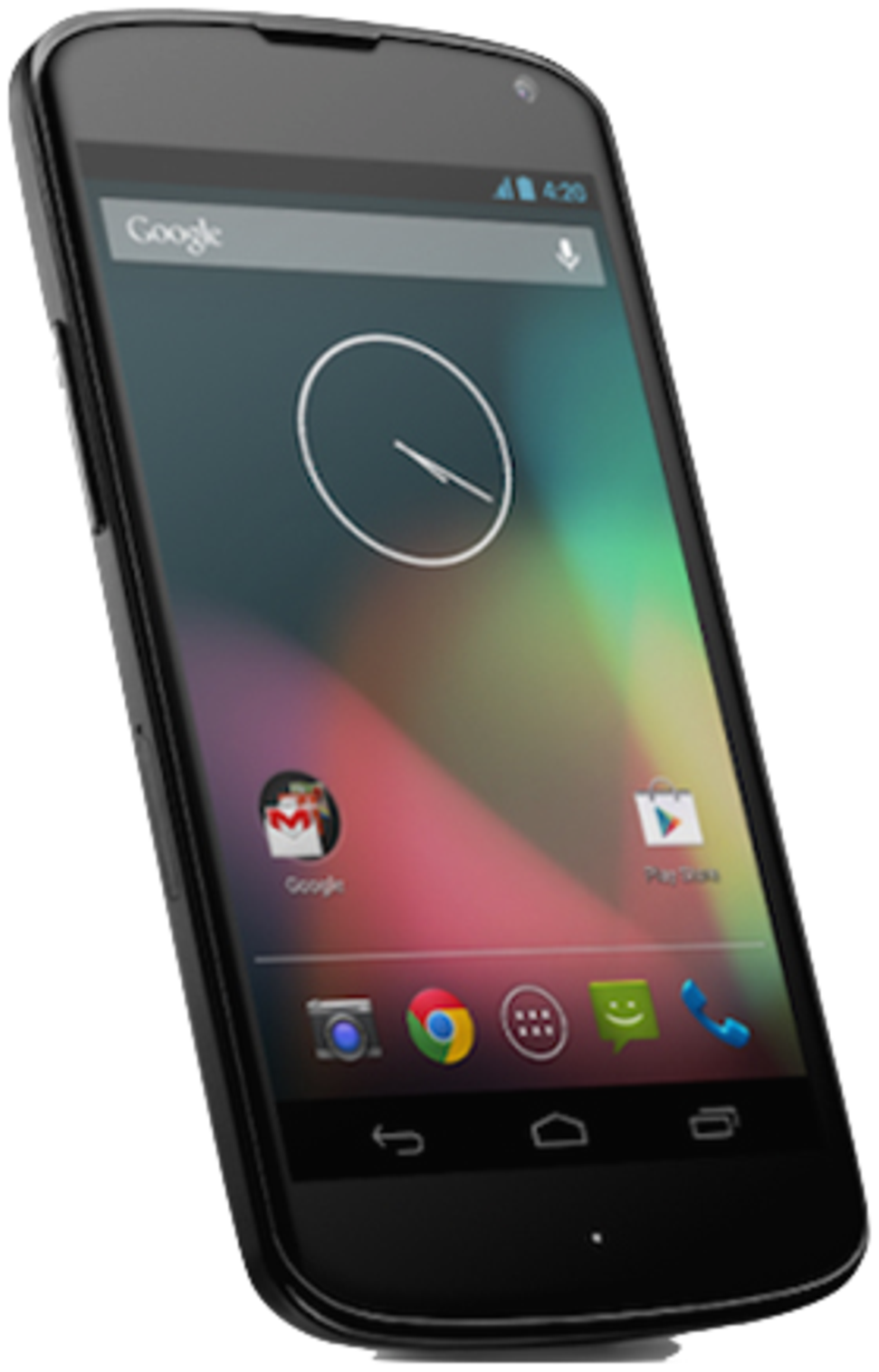}
      \\Nexus 4
      \end{minipage}\\
\hline \hline
      \multirow{3}{*}{
      \begin{minipage}{.07\textwidth}
      \includegraphics[height=8mm]{samsung_3}
      \end{minipage}}
      &\textbf{Avg}&$0.62$ & $0.85$ & $3.31$ \\
      &\textbf{50\%}& $0$ & $0$ & $2.83$ \\
      &\textbf{75\%}& $0$ & $2$ & $4.47$ \\ \hline

      \multirow{3}{*}{
      \begin{minipage}{.07\textwidth}
      \includegraphics[height=8mm]{z2_3}
      \end{minipage}}
      &\textbf{Avg} &$1.48$ & $0.27$ & $1.8$ \\
      &\textbf{50\%}&$0$ & $0$ & $0$ \\
      &\textbf{75\%}&$2.83$ & $0$ & $2.83$\\ \hline

       \multirow{3}{*}{
      \begin{minipage}{.07\textwidth}
      \includegraphics[height=8mm]{nexus4}
      \end{minipage}}
      &\textbf{Avg}&$2.8$ & $2.15$ & $1.99$ \\
      &\textbf{50\%}&$2$ & $2$ & $2$ \\
      &\textbf{75\%}&$4$ & $2$ & $2$\\ \hline
  \end{tabular}
\caption{Localization error using different pairs of phones for training and testing. 
 Each cell contains three values: mean, median, and 75th percentile of localization error.}
    \label{tab:diff_mobiles_loc_erro}
    \vspace{-0.2in}
\end{table}

\subsection{Case Study}
\label{sec:case_study}
In this section, we extensively evaluate the performance of \sys{} using the calculator application shown in Figure~\ref{fig:real_kids_calculator}. For the evaluation purpose, four users used the application over seven days (different magnets/devices) and the user counted the number of mis-detections (if any). 
The results show that \sys{} has 100\% key detection accuracy. This can be explained by noting that all cells in the calculator area (The red rectangle in Figure~\ref{fig:heatmap}) have a localization error less than one centimetre. In addition, the size of the smallest calculator button is 4cm$\times$4cm. Note that larger keyboards may cover the entire grid area and can lead to lower accuracy than in this particular case study. However, \sys{} can still provide its 100\% accuracy in the majority of cases as discussed in the next section.

We also tested the calculator application on Samsung mini S4, Sony Z2, and Sony Xperia tipo with standard touch screen. The number of covered buttons in the experiment is 16, each button is clicked 20 times. The accuracy for both Samsung mini S4 and Sony Z2 is around 99\% (3 misses for Samsung mini S4 and 2 misses for Sony Z2 out of 320) and the accuracy for Sony tipo is 98\% (5 misses out of 320).

\section{Discussion}
\label{sec:discussion}
\sys{} provides a ubiquitous homomorphic keyboard that can suit different users' needs. All that is required is a phone with a magnetometer and an off-the-shelf magnet, even those available in children toys. \sys{}'s ubiquity comes from its independence from the background magnetic interference, phone orientation, polarity, devices heterogeneity, and differences between magnets. The independence of magnetic interference and phone orientation comes from the offset removal process. The first-time use calibration process that maps the factory calibrated fingerprint to a new phone/magnet makes it independent of the devices heterogeneity and magnets differences. Finally, changes in polarity are handled through a simple comparison with a reference reading as part of enabling the system.

\sys{} currently supports discrete detection of pressed keys. However, there are a number of applications that require continuous tracking, such as a virtual mouse, touch pads, paint programs, and some games. Our initial results on localization accuracy in Section~\ref{sec:impact_of_spatial_averaging_parameter_k} shows promising tracking results in the continuous case of less than 10mm, which can be further improved algorithmically. Due to space constraints, we leave this to future work.

Currently, \sys{} is designed to track a single magnet. This is useful for a large number of applications, e.g. applications for kids and the visually impaired as well as for the continuous tracking case as in drawing applications. However, supporting multiple magnet tracking can be useful for a number of applications, such as full keyboard typing. Multi-object tacking is a well-studied problem in other domains (e.g. computer vision \cite{krumm2000multi} and device-free localization \cite{sabek2012multi}), where techniques can be borrowed. We leave the multi-magnet localization problem for a future paper.  

Noting that the further away from the magnet the lower the accuracy of detection due to the weak magnetic field, the application designer can take this into account in the ``\emph{Keyboard Designer}'' module, e.g. by placing less frequently used keys at locations with higher errors. This can also be automated by a new system module.

Finally, co-occurrence models of different keys can be used to further enhance accuracy by dynamically changing the new key probabilities based on the previous detected keys. This can incorporated easily in our probabilistic estimation framework as $P(x)$ in Equation~\ref{equ:bayesian_inversion}.

\section{Related Work}
\label{related_work}
In this section, we discuss various related work to \sys{} in two areas: magnet-based interaction systems and keystrokes recognition systems.

\subsection{Magnet-based Interaction Systems}
The human body is transparent to static and low-frequency magnetic fields, which makes magnetic tracking a promising technique in a number of human-computer interaction applications. A number of systems have been proposed for the continuous tracking of a magnet location based on nonlinear models of the relation between the magnetic field and distance \cite{hu2005efficient,han2007wearable}. 
However, such systems usually require multiple external high-accuracy magnets and/or assume that the magnet diameter is very small relative to the distance between the magnet and the point (\emph{limiting the tracking area to a few millimeters}). \sys{}, on the other hand, can be operate with magnets with any size and provides millimeter-level accuracy for much larger areas. Similarly, MagPen \cite{hwang2013magpen} uses a \textbf{\emph{specially designed hardware}} stylus that combines a magnet with capacitive sensing to extend the functionality of the current styluses.

Abracadabra \cite{harrison2009abracadabra} is a magnetically-driven input approach for smart watches. It senses the movement of a magnet placed on the fingertip to determine its relative position from the watch. This information is used to control a cursor, make selections, or issue basic gestures. Abracadabra, though, depends on a high-accuracy external magnetometer sensor to avoid the noises inside the mobile device. Similarly, Nenya \cite{ashbrook2011nenya} employs an external magnetometer to track two specific motions of ring: twisting around the user finger and sliding along the finger. Both Abracadabra and Nenya have been evaluated with \textbf{\emph{a specific phone and magnet}}. \sys{}, in contrast, uses standard phone sensors, which have higher noises, and can work with any phone or magnet as we show and quantify in the paper. In addition, it can differentiate between a large number of cells and, hence, is can be used with a wider set of applications. 

The MagiWrite \cite{ketabdar2010magiwrite} and MagSign \cite{ketabdar2010magisign} systems use a machine learning approach to detect few specific patterns of the magnetometer motion. Specifically, MagiWrite detects only the \textbf{\emph{10 digits}} and MagSign detects the \textbf{\emph{one specific}} user signature for authentication purposes.
\sys{}, on the other hand, is designed to work with any keyboard and can differentiate a large number of cells. In addition, it has been tested with different phones and magnets and can handle different magnet polarities.

\sys{} presents a homomorphic ubiquitous keyboard based on magnetic field. Its main advantage is the ability to work with any mobile phone or magnet with minimal effort. In addition, it gives both the application developers and users the flexibility of designing their custom keyboards that fit different needs. 

\subsection{Keystrokes Recognition Systems}
A number of techniques have been proposed in literature to recognize keystrokes including vision-based \cite{balzarotti2008clearshot}, acoustic-based \cite{zhu2014context,asonov2004keyboard,wang2014ubiquitous}, electromagnetic emission-based \cite{vuagnoux2009compromising}, and WiFi-based approaches \cite{ali2015wikey}. Vision based approach, e.g. \cite{balzarotti2008clearshot}, recognizes keystrokes based on analyzing the captured video of the users' fingers. Acoustic-based approach recognizes keystrokes either based on the time the sound takes from the pressed key to the acoustic receiver, e.g. \cite{zhu2014context}, or based on the different acoustic signatures of the different keys on the keyboard, e.g. \cite{asonov2004keyboard, wang2014ubiquitous}.  On the other hand, electromagnetic emission-based approaches, e.g. \cite{vuagnoux2009compromising}, recognize keystrokes based on that the electromagnetic emission from the electric circuits underneath different keys in a keyboard. 
WiFi-based approaches extend the concept of WiFi-sensing \cite{abdelnasser2015ubibreathe}, \cite{abdelnasser2015wigest} and device free localization \cite{seifeldin2010deterministic}, \cite{el2010propagation}, \cite{eleryan2011aroma}, \cite{kassem2012rf}, \cite{saeed2014ichnaea}, \cite{abdel2013monophy} and leverage the channel state information of the received MIMO WiFi system to detect the finger location \cite{ali2015wikey}. All these systems suffer from occlusion; interference from other movement, humans, surrounding noises; and/or high energy consumption. In addition, they are only tested with specific devices.

In contrast, \sys{} proposes a novel technique of keystrokes recognition based on magnetic field. The magnetic field is a more promising technique as it is transparent to human motion and the magnetometer consumes order of magnitude less energy compared, e.g., to the camera or WiFi. Moreover, \sys{} has different modules to handle the magnetic interference and designed and tested to work with any mobile device or magnets.

\section{Conclusion}
\label{sec:conclusion}
In this paper, we have shared the design and implementation of \sys{}. \sys{} leverages a one-time factory-based fingerprint in a probabilistic framework to enable custom-designed keyboards that can fit different user's needs. It has a number of modules that makes it work with heterogeneous devices and magnets with different shapes, sizes, and strengths. In addition, it can handle other practical deployment scenarios such as changes in magnet polarity.

Evaluation of \sys{} on different Android devices in various realistic scenarios shows that it can achieve 91\% key detection accuracy for keys as small as 2cm$\times$2cm, increasing to 100\% for  4cm$\times$4cm keys.  This accuracy is robust to different phones and magnets. In addition, we presented a detailed evaluation of a custom calculator case study showing a detection accuracy of $100\%$ for all keys.

\bibliographystyle{abbrv}

\end{document}